\documentclass[english,pra,twocolumn,superscriptaddress, eqsecnum, floatfix]{revtex4}
\usepackage[T1]{fontenc}
\usepackage[latin9]{inputenc}
\usepackage{color}
\usepackage{float}
\usepackage{amsmath}
\usepackage{graphicx}
\usepackage{amssymb}

\makeatletter
\@ifundefined{textcolor}{}
{%
 \definecolor{BLACK}{gray}{0}
 \definecolor{WHITE}{gray}{1}
 \definecolor{RED}{rgb}{1,0,0}
 \definecolor{GREEN}{rgb}{0,1,0}
 \definecolor{BLUE}{rgb}{0,0,1}
 \definecolor{CYAN}{cmyk}{1,0,0,0}
 \definecolor{MAGENTA}{cmyk}{0,1,0,0}
 \definecolor{YELLOW}{cmyk}{0,0,1,0}
 }

\makeatother

\usepackage{babel}

\begin{document}

\title{Bell inequalities for Continuous-Variable Measurements }

\author{Q. Y. He}

\affiliation{ARC Centre of Excellence for Quantum-Atom Optics, Swinburne University
of Technology, Melbourne, Australia.}

\author{E. G. Cavalcanti}

\affiliation{Centre for Quantum Dynamics, Griffith University, Brisbane QLD 4111,
Australia}

\author{M. D. Reid}

\affiliation{ARC Centre of Excellence for Quantum-Atom Optics, Swinburne University
of Technology, Melbourne, Australia.}

\author{P. D. Drummond}

\affiliation{ARC Centre of Excellence for Quantum-Atom Optics, Swinburne University
of Technology, Melbourne, Australia.}
\begin{abstract}
Tests of local hidden variable theories using measurements with continuous
variable (CV) outcomes are developed, and a comparison of different
methods is presented. As examples, we focus on multipartite entangled
GHZ and cluster states. We suggest a physical process that produces
the states proposed here, and investigate experiments both with and
without binning of the continuous variable. In the former case, the
Mermin-Klyshko inequalities can be used directly. For unbinned outcomes,
the moment-based CFRD inequalities are extended to\emph{ }functional
inequalities by considering arbitrary functions of the measurements
at each site.\textcolor{black}{{} By optimising these functions, we
obtain more robust violations of local }hidden variable theories\textcolor{black}{{}
than with either binning or moments. }Recent inequalities based on
the algebra of quaternions and octonions are compared with these methods.
Since the prime advantage of CV experiments is to provide a route
to highly efficient detection via homodyne measurements, we analyse
the effect of noise and detection losses in both binned and unbinned
cases. \textcolor{black}{The CV moment inequalities with an optimal
function have greater robustness to both loss and noise. This could
permit a loophole-free test of Bell inequalities.} 
\end{abstract}
\maketitle

\section{INTRODUCTION}

In 1964 Bell showed that the predictions of quantum mechanics are
not compatible with those of local hidden variable theories (LHV)
\cite{Bell}. Since then, there have been experimental tests \cite{CHSH,aspect,Zeilinger1999}
of these predictions, which can be used to rule out some LHV alternatives
to quantum mechanics (QM). While experiments to date support quantum
mechanics, the experimental strategies used have loopholes \cite{loophole}.
In some experiments, problems are caused by insufficient spatial separation
of measurements \cite{Winelandiontrap}---so that the assumption of
local causality that underpins the LHV model cannot be applied to
start with. In other experiments, the poor photon detection efficiencies
\cite{ZeilingerBell} mean that correlations are not strong enough
to violate a Bell inequality, unless additional assumptions are utilized.
In short, a clear violation applicable to all local hidden variable
theories has not yet been experimentally observed. 

It has been widely suggested that the use of highly efficient detection
techniques like homodyne detection \cite{cvbin} may provide an avenue
to overcome the detection efficiency problems \cite{loophole} with
Bell's theorem. Already, continuous variable (CV) realizations of
the Einstein-Podolsky-Rosen (EPR) paradox have been achieved with
high efficiencies \cite{OuKimbleEPR}, using optical homodyne detection
\cite{rmpepr}. In this case, the EPR paradox is reformulated in terms
of field quadrature phase amplitudes which have commutators like position
and momentum. The EPR paradox, however, only demonstrates the incompatibility
of local realism (LR) with the completeness of quantum mechanics.
It does not test local realism itself. Proposals to demonstrate an
incompatibility of quantum theory with local hidden variable theories
using continuous variables require new formalisms, since Bell's original
inequality was only applicable to discrete or binned outcomes.

Early proposals for such tests using CV measurements \cite{cvbin}
relied on binning the initially continuous outcome domains to obtain
binary results. Here the original Bell Clauser-Horne-Shimony-Holt
inequalities \cite{CHSH-1} are applicable. However, these predicted
violations of Bell's inequality were small and involved states that
were difficult to prepare. Wenger \emph{et al.} advocated a root binning
technique to enhance the level of violation \cite{wenger}. Nha and
Carmichael \cite{NhaCarmichael} proposed a conditional state preparation
to obtain the required nonclassical states. Munro \emph{et al. }\cite{munro}
and Acin \emph{et al.} \cite{Acin } suggested using binned CV measurements
made on $N$ spatially separated qubit systems prepared in a Greenberger-Horne-Zeilinger
(GHZ) state \cite{mermin}, and noted improved violation of the Mermin-Ardehali-Belinskii-Klyshko
(MABK) Bell inequalities \cite{Klyshko} as $N$ is increased. 

More recently, Cavalcanti\emph{, }Foster, Reid and Drummond\emph{
}(CFRD) \cite{cvbell2} have derived a continuous variable inequality
for LHV theories, formulated directly in terms of moments, without
binning. This allows a simple test of local hidden variable theories
for genuine CV outcomes. Violations of this CV moment inequality were
found possible for GHZ states involving a large number of sites ($N\geq10$).
Based on the CFRD inequalities, \textcolor{black}{Shchukin and Vogel$\ $\cite{Shchukin}
introduced further inequalities considering $4$ or $8$ observables
at each site. Salles et al have shown\cite{SallesCav} that at least
three sites are necessary for the CFRD inequalities to be violated,
and have more recently generalized this technique to multilinear mappings\cite{SallesCav2}.
In general, the use of ordinary moments has the advantage that it
is more robust against loss and noise than earlier binning or conditional
techniques. However, the large number of multipartite sites involved
presents practical difficulties.}

In this paper, we develop \emph{functional} CFRD Bell inequalities
for testing local hidden variable theories, using arbitrary functions
of continuous variable outcomes at each site. Variational calculus
is used to optimize the choice of measured function, to maximise the
violation of the Bell inequality. As a result, we obtain predictions
for violations of LHV models with fewer sites, larger losses and greater
degrees of decoherence than previously possible. This approach of
optimising the function of a measurement result for a particular purpose
should be generally useful in the field of quantum information. Preliminary
results have been presented as a short Letter \cite{functional letter}.
Here we provide a more detailed and thorough account of this procedure.

The work of Acin \emph{et al.} \cite{Acin } and Cavalcanti \emph{et
al. }\cite{cvbell2} has revealed an exponential improvement in Bell
inequality violations as the number of sites $N$ increases, for CV
measurements. These results support the earlier work of MABK, which
also showed an exponential improvement of violation for $N$-partite
states for spin measurements. Since these inequalities require detection
at every site, one might expect this improvement to be counteracted
by increasing sensitivity to detection losses. Work on the robustness
of Mermin-Bell and CFRD-Bell violations with loss \cite{lossN,cvbell2}
shows that this is not the case. Other calculations using different
LHV assumptions \cite{recent prl crit eff-1} have suggested a loss
threshold as low as $\eta_{crit}=0.5$ as $N\rightarrow\infty$. However,
since this proposal uses additional assumptions about the LHV models,
it may not rule out all LHV theories. 

In the present paper, we introduce a new approach to nonlocality that
utilizes a general functional optimization of continuous variable
observables.\textcolor{black}{{} }We~\textcolor{black}{find the optimal
function that maximises a violation of the CV Bell inequality for
a given efficiency $\eta$ and state purity $p$. We show that the
optimal function has the form }\textcolor{black}{\emph{$x/(1+\varepsilon_{N}x^{2})$,}}\textcolor{black}{{}
where $\varepsilon_{N}$ is a parameter related to $N$ and $\eta$.
This gives an inequality for all LHV models, which is violated by
the GHZ states of (\ref{eq:state}) for $N\geq5$. The violation increases
exponentially with $N$. For a pure state the loss threshold $\eta_{crit}$
decreases asymptotically to $\eta_{\infty}=0.69$ as $N\rightarrow\infty$,
}in the most symmetric case\textcolor{black}{. Thus} both the number
of modes required, and the required efficiency are dramatically reduced.
The wider range of quantum states considered here include states 
which can be straightforwardly obtained from a polarisation GHZ state
of the type generated in current experiments. These are different
to the extreme photon-number correlated states considered by Acin
\emph{et al}, which may be an experimental advantage. 

A simple example of a functional transformation is obtained when binning
of a CV observable is used to give discrete binary outcomes \cite{cvbin}.
In this case, the CFRD inequalities reduce to those of Mermin \cite{mermin}.
Here, our analysis is similar to that of Acin \emph{et al}. \cite{Acin }.
For binary outcomes, one can use extensions of Mermin's inequalities
\cite{Klyshko} for the extreme \textcolor{black}{photon-number correlated}
states $\left|N\right\rangle =\left(|0\rangle^{\otimes N}+|1\rangle^{\otimes N}\right)/\sqrt{2}$.
One then obtains an exponential increase of violation of LHV with
$N$, and violation for $N\geq3$. In addition, we extend the analysis
of Mermin and Acin \emph{et al}. \cite{Acin }, and calculate results
for homodyne detection for more feasible types of state. However,
we find $(\eta p^{2}){}_{crit}=2^{(1-2N)/N}\pi$, which means that
this approach is more susceptible to loss and decoherence than using
functional moment inequalities. The critical efficiency for a pure
state at large $N$ is $\eta=0.79$, which is substantially higher
than for the optimal functional CFRD approach.

The organisation of the paper is as follows. Section II gives the
general notation used for continuous variable moments, and reviews
the original CFRD inequality. In Section III we review and extend
the CFRD inequality for the case of quadrature measurements on generalized
GHZ states. Section IV considers the multipartite binning approach
that uses the MABK Bell inequalities, which was also studied recently
by Acin \emph{et al}. for quadrature detection. In both Sections III
and IV, we consider several types of GHZ state, and also present results
for cluster and W-states \cite{wstate,cluster}. Next, in Section
V these approaches are generalized to include functions of the observables.
We compute the maximum violation of the inequality for several choices
of function, in the case of a generalized multipartite GHZ state.
Functional transformations using an optimum function, i.e., \textcolor{black}{\emph{$x/(1+\varepsilon_{N}x^{2})$,}}
are shown to give a substantial improvement to results obtained with
the CFRD approach. In Section VI these results are compared with the
Shchukin-Vogel (SV) inequality for larger numbers of observables,
showing that this gives identical results to the simpler CFRD method,
in the cases studied. In Section VII we include the effects of loss
and noise on the violations of the functional moment inequalities
and MABK inequalities. Optimizing the choice of measured function
reveals a quantum nonlocality for larger losses and for greater degrees
of decoherence than possible previously. Section VIII gives an overview
of the results.

In summary, we examine in this paper the effect of loss and state
preparation impurity on tests of local realism using functional moments.
Our motivation is the fundamental challenge of developing an experimentally
feasible loop-hole free test of local realism, which balances the
need for a simple experimental realization with robustness against
loss and noise.

\section{Continuous variable measurements}

Early work on Bell inequalities utilized measurements of spin, making
it seem that obtaining discrete measurement outcomes was a crucial
ingredient to violating a Bell inequality. In fact, this is not essential.
The CFRD inequality is a multipartite test of local realism involving
ANY two observables measured at each of $N$ sites, with causal separation
between measurement at each site.

\subsection{General CFRD inequality}

It was shown in \cite{cvbell2} that the following inequality must
hold for all LHV theories: \begin{equation}
\left|\left\langle \prod_{n=1}^{N}(x_{n}+ip_{n})\right\rangle _{P}\right|^{2}\leq\left\langle \prod_{n=1}^{N}(x_{n}^{2}+p_{n}^{2})\right\rangle _{P},\label{eq:CFRD}\end{equation}
where $x_{n}$, $p_{n}$ are the outcomes of two arbitrary measurements
\cite{key-1} at each site labeled $n$. The measurements are causally
separated, and $\left\langle \right\rangle _{P}$ indicates an average
over a set of local hidden variables $\lambda$ with probability $P\left(\lambda\right)$. 

In quantum mechanics, these correlations involve operators $\hat{x}_{n}$,
$\hat{p}_{n}$ at each site $n$, with eigenvalues $x_{n},\,\, p_{n}$.
Unlike in hidden variable theories, these inequalities \emph{can}
be violated for certain quantum states, if the operators do not commute.
Despite the notation used, nothing is assumed about the type of measurement
operators. In LHV theories, the inequality must be satisfied by measurements
with any type of spectrum, discrete or continuous, and even unbounded
outcomes. For this inequality to be useful, one must find states and
observables that violate the inequality within quantum theory, and
are obtainable in an experiment. 

It is important to recall here that when $\hat{x}_{n}$, $\hat{p}_{n}$
are described within quantum mechanics by non-commuting operators,
the corresponding physical variables are not simultaneously measurable.
Thus, the $N$-fold products in the above inequality each involve
a sum over $2^{N}$ distinct correlation measurements with different
settings of the measurement apparata.

\subsection{Quadrature measurements}

While the inequality is valid in general, we wish to apply it to experiments
in quantum or atom optics, which are known to be able to produce highly
correlated, nonclassical quantum states. Operationally, we will consider
the case where conventional photodetectors are replaced by a homodyne
detector using a local oscillator, in order to detect field quadratures
rather than photon numbers. As well as involving an intrinsically
efficient experimental technique, this method is already known to
be robust against loss, without additional assumptions. Although quadrature
measurements are traditionally used for coherent and squeezed states,
they of course can also be used for states with well-defined overall
particle number. In principle, exactly the same techniques apply to
any boson fields, including, for example, ultra-cold atoms.

We assume there are $N$ sites at which one can make observations,
with an adjustable local oscillator phase to define local quadrature
operators $\hat{x}_{n}$, $\hat{p}_{n}$. In the case that $\hat{x}_{n}$
and $\hat{p}_{n}$ are associated with single-mode bosonic creation
and annihilation operators $\hat{a}_{n}^{\dagger},$ $\hat{a}_{n}$,
one has $\hat{x}_{n}=\left(\hat{a}_{n}+\hat{a}_{n}^{\dagger}\right)/2$
and $\hat{p}_{n}=i(\hat{a}_{n}^{\dagger}-\hat{a}_{n})/2$. These are
the quadrature phase amplitudes associated with the boson annihilation
operator $\hat{a}_{n}$, or for suitable choice of units, (see the
Appendix) the position and momentum of a harmonic oscillator, with
commutators: \begin{equation}
\left[\hat{x}_{m},\hat{p}_{n}\right]=\frac{i}{2}\delta_{mn}\,\,.\end{equation}
One can adjust the local oscillator phase to detect linear combinations
of position and momentum. Hence, we can introduce a general quadrature
phase operator: \begin{eqnarray*}
\hat{X}_{n}^{\theta} & = & \hat{x}_{n}\cos\theta+\hat{p}_{n}\sin\theta\\
 & = & \frac{1}{2}\left(\hat{a}_{n}e^{-i\theta}+\hat{a}_{n}^{\dagger}e^{i\theta}\right)\ .\end{eqnarray*}

For continuous measurements of quadrature variables on entangled boson
states, we define generalized GHZ states to be:\begin{equation}
\left|\psi\right\rangle =\left(|0\rangle^{\otimes r}|1\rangle^{\otimes N-r}+|1\rangle^{\otimes r}|0\rangle^{\otimes N-r}\right)/\sqrt{2}\,\,.\label{eq:state}\end{equation}
Here the integer $r$ is the maximum number of modes having nonzero
photon number. The critical efficiency $\eta_{crit}$ required for
violation of the CFRD inequality in the symmetric case of $r=N/2$
tends to $\eta_{crit}=0.81$, as $N\rightarrow\infty$. Quadrature
measurements with local oscillators are inherently highly efficient
compared to spin or photon-number measurements, with reported efficiencies
of $99\%$. However, generation losses from mode-matching can degrade
the experimental efficiency, so $81\%$ is still a challenging practical
benchmark, especially with large numbers of correlated modes.

\subsection{Functional transformations}

Functional transformations of any output measurements are always possible
experimentally. Any data file of experimental results is trivially
processed into a function or set of functions of the original measured
data. While this procedure is not very useful for binary results,
it represents a large functional space of measurements in the continuous
variable case.

In terms of hidden-variable theories, this simply means that all measured
quantities $x$ can be replaced by a function $f\left(x\right)$.
In quantum mechanics, this implies that all eigenvalues $e$ are replaced
by $f\left(e\right)$, or in operator language, the measurement operator
$\hat{x}$ is replaced by $f\left(\hat{x}\right).$ The details of
this procedure are explained in later sections. We will show that
such functional transformations greatly reduce the number of sites
and the critical efficiencies needed to violate a Bell inequality.

\section{CFRD inequality }

In this section we will review the CFRD inequality \cite{cvbell2}
as applied to quadrature measurements, and provide a more complete
analysis of its violation than in the Letter that introduced it. Quadrature
observables at $N$ sites are considered here, as described in the
previous section.

\subsection{Quadrature CFRD inequality}

At each site, the sign associated with $\hat{p}_{n}$ can be chosen
to obtain either $\hat{x}_{n}+i\hat{p}_{n}=\hat{a}_{n}$ or $\hat{x}_{n}-i\hat{p}_{n}=\hat{a}_{n}^{\dagger}$
for each of the terms in the left hand side ($LHS$) of \eqref{eq:CFRD}.
The choice of sign is important to maximise the violations for a given
state. Defining a variable $s_{n}\in\{-1,1\}$ to represent this choice
for each site, we denote by $A_{n}(-1)=a_{n}$ and $A_{n}(1)=a_{n}^{\dagger}$
the corresponding combination of measurements at each site. The CFRD
inequality then reduces to \begin{equation}
\left|\left\langle \prod_{n=1}^{N}A_{n}(s_{n})\right\rangle _{P}\right|^{2}\leq\left\langle \prod_{n=1}^{N}\left(x_{n}^{2}+p_{n}^{2}\right)\right\rangle _{P}\,.\label{eq:CFRD_QM}\end{equation}
CFRD showed that within quantum theory, the symmetric state \begin{equation}
\left|\frac{N}{2}\right\rangle =\frac{1}{\sqrt{2}}\left(|0\rangle^{\otimes N/2}|1\rangle^{\otimes N/2}+|1\rangle^{\otimes N/2}|0\rangle^{\otimes N/2}\right)\label{eq:N/2 state}\end{equation}
violates this inequality for $N\geq10$. \textcolor{black}{Note here
that there is a spatial mode index that has been dropped for convenience
but is given by the order of the kets}, so that this state represents
$N$ distinct modes at spatially separated sites, and each mode is
occupied by one or zero photons. Thus, the states $|0\rangle$, $|1\rangle$
are eigenstates of the number operator $a^{\dagger}a$ with corresponding
eigenvalues $0$ and $1$ respectively, so the prediction could in
principle be tested with photonic GHZ states produced in the laboratory.
While a value of $N=10$ is not impossible, it is a large number of
correlated modes for a practical experiment which involves multiple
down-conversions and beam-splitters. 

\begin{figure}[h]
\textcolor{black}{\includegraphics[width=0.9\columnwidth]{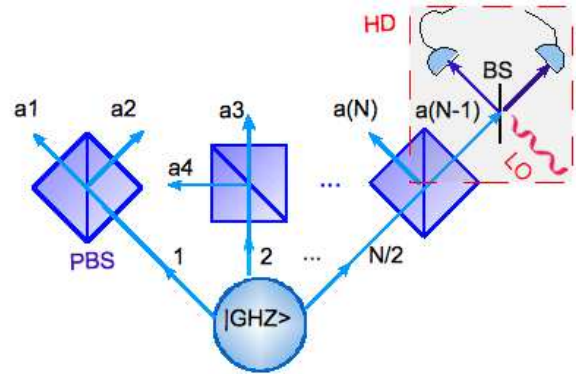}}

\textcolor{black}{\caption{\textcolor{black}{(Color online) Schematic of an experimental setup
to test the CFRD inequalities. We assume an $N/2$-photon GHZ state
can be generated}.\textcolor{black}{{} Then by using a PBS, each photon
is sent to one of the two output spatial modes depending on its polarisation.
The entanglement in polarization is therefore translated into entanglement
in spatial modes to }produce the symmetric states $\left|N/2\right\rangle $
with\textcolor{black}{{} $\ N$ field modes. At each site, high-efficiency
homodyne detection is used to measure the correlations.\label{fig:Schematic}}}
}
\end{figure}

This type of symmetric GHZ state is obtainable from the usual GHZ
polarized state via a local unitary transformation at each location.
For example, the symmetric state given above can be prepared from
an $N/2$-photon polarization GHZ state, of a type that is found in
some current experiments \cite{expghz,clusterexpfour-photon,six-photon}
\begin{equation}
\left|GHZP\right\rangle =\frac{1}{\sqrt{2}}\left(|H\rangle^{\otimes N/2}+|V\rangle^{\otimes N/2}\right)\,.\end{equation}
A possible procedure is shown in Fig.~\ref{fig:Schematic}, where
$|H\rangle,\ |V\rangle$ represent horizontally or vertically polarized
single-photon states, by passing each photon through a polarizing
beam splitter (PBS). This converts polarization modes into spatial
modes, so that $|H\rangle\equiv|1\rangle_{H}|0\rangle_{V}$ and $|V\rangle\equiv|0\rangle_{H}|1\rangle_{V}$
where $|0\rangle_{H/V}$ and $|1\rangle_{H/V}$ are number states
for the horizontally and vertically polarised modes respectively.
The $|GHZP\rangle$ state then becomes $|\frac{N}{2}\rangle$ of (\ref{eq:N/2 state}). 

On-demand state generation

Some caution is needed here in interpreting current GHZ experiments
\cite{expghz}. These typically involve conditional measurements,
rather than a known initial quantum state. To our knowledge, current
experiments only generate GHZ states in the sense that state is identified
after some number of photon-counting events have already taken place.
Here we assume a more traditional quantum mechanical state preparation,
in that we assume a GHZ state can be generated on demand. We note
that a proposal already exists for carrying out on-demand generation
of a singlet state\cite{Sliwa}, which is the simplest case. Current
experimental techniques for GHZ states would therefore need similar
modifications to produce a well-defined GHZ state `on demand'. 

\textcolor{black}{}%

\subsection{Generalized GHZ states}

\textcolor{black}{A class of states that violate the CFRD inequalities
is represented by the entangled states \begin{equation}
|\psi\rangle=c_{1}|0\rangle^{\otimes r}|1\rangle^{\otimes(N-r)}+c_{2}|1\rangle^{\otimes r}|0\rangle^{\otimes(N-r)}\ .\label{eq:GHZ}\end{equation}
Thus $r=N$ corresponds to extreme photon-number correlated states},
\textcolor{black}{a superposition of a state with $0$ photons at
all sites and a state with $1$ photon at each site, while other values
of $r$ correspond to intermediate cases. These can of course be transformed
into each other with local unitary operations, although in practical
terms such transformations are not always feasible without losses.}

Within quantum mechanics, the $LHS$ of the CFRD inequality \eqref{eq:CFRD_QM}
is non-zero with the choice \begin{equation}
LHS=\left|\left\langle \hat{a}_{1}^{\dagger}..\hat{a}_{r}^{\dagger}\hat{a}_{r+1}...\hat{a}_{N}\right\rangle \right|^{2}=\mid c_{1}c_{2}\mid^{2}\ .\label{eq:LHS}\end{equation}
\textcolor{black}{Here, $|c_{1}|^{2}+|c_{2}|^{2}=1$, and the violation
is} maximized for\textcolor{black}{{} $\ c_{1}=c_{2}=1/\sqrt{2}$}.
For the right hand side ($RHS$) of the inequality, we obtain \textcolor{black}{\begin{eqnarray}
RHS & = & \langle(\hat{a}_{1}^{\dagger}\hat{a}_{1}+1/2)...(\hat{a}_{N}^{\dagger}\hat{a}_{N}+1/2)\rangle\nonumber \\
 & = & c_{1}^{2}(1/2)^{r}(3/2)^{N-r}+c_{2}^{2}(1/2)^{N-r}(3/2)^{r}.\label{eq:RHSvalue}\end{eqnarray}
}If we use photon-number correlated states with $r=N/2$, \begin{equation}
RHS=(\frac{3}{4})^{N/2}\ ,\end{equation}
\textcolor{black}{which is independent of the amplitudes $c_{1},\ c_{2}$}. 

Alternately, if we \textcolor{black}{fix $c_{1}=c_{2}=1/\sqrt{2}$
but change $r$, }

\begin{equation}
RHS=\frac{3^{N-r}+3^{r}}{2^{N+1}}\ .\label{eq:RHS}\end{equation}

\begin{figure}[H]
\textcolor{black}{\includegraphics[width=1\columnwidth]{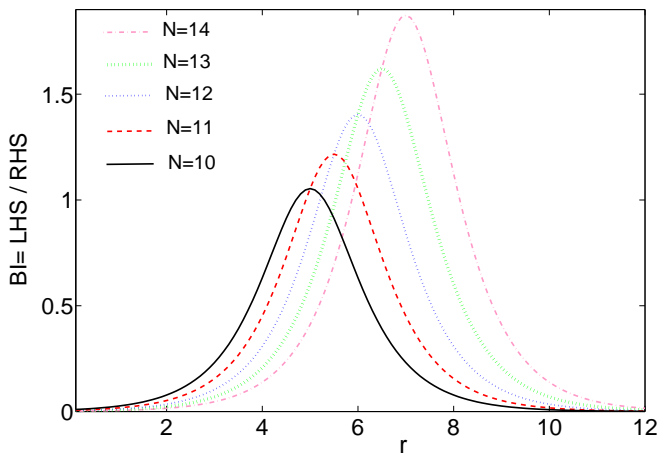}}

\textcolor{black}{\caption{\textcolor{black}{(Color online) Bell violation }as a function of
$r$ with $c_{1}=c_{2}=1/\sqrt{2}$. Different $N$ (10-14) values
are distinguished by their line style. The allowed values of \emph{r
}are of course only the natural numbers; the lines are drawn to guide
the eye.\label{fig:Bell with r}}
}
\end{figure}

There is a violation of the inequality when the Bell observable $B_{N}=LHS/RHS>1$.
We note from Fig. \ref{fig:Bell with r} that no violations are possible
for $r=N$, and the optimal case has $r$ approximately equal to $N/2$.
For $N$ even, where $r=N/2$ there is a violation if

\begin{equation}
B_{N}=\frac{1}{4}(\frac{4}{3})^{\frac{N}{2}}>1\end{equation}
in agreement with Cavalcanti \emph{et al.} \cite{cvbell2} who have
shown this requires $N\geq10$. For odd $N$, we select $r=(N\pm1)/2$
to obtain a violation if\begin{equation}
2^{N-1}>3^{(N-1)/2}+3^{(N+1)/2}\ ,\end{equation}
which is obtained for $N\geq11$.

\textcolor{black}{}%

\subsection{\textcolor{black}{Cluster and W states}}

\textcolor{black}{Other forms of multipartite entangled states, such
as W-states and cluster states \cite{wstate,cluster}, are also of
interest, either because they are simpler to generate experimentally,
or are more resistant to certain forms of decoherence. In addition
to the GHZ states \cite{expghz}, four and six-photon cluster states
have been realized experimentally \cite{clusterexpfour-photon,six-photon}.
We thus present initial results for the CFRD inequality using cluster
and W-states. }

\textcolor{black}{Consider first the four-qubit cluster state \cite{cluster,clusterexpfour-photon,six-photon}
defined as }

\textcolor{black}{\begin{eqnarray}
|C\rangle & = & \frac{1}{2}(|1\rangle|1\rangle|1\rangle|1\rangle+|1\rangle|1\rangle|0\rangle|0\rangle\nonumber \\
 &  & +|1\rangle|1\rangle|0\rangle|0\rangle-|0\rangle|0\rangle|0\rangle|0\rangle)\label{eq:cluster state}\end{eqnarray}
and also consider the four-qubit W-state \cite{wstate}}

\textcolor{black}{\begin{eqnarray}
|W\rangle & = & \frac{1}{2}(|1\rangle|0\rangle|0\rangle|0\rangle+|0\rangle|1\rangle|0\rangle|0\rangle\nonumber \\
 &  & +|0\rangle|0\rangle|1\rangle|0\rangle+|0\rangle|0\rangle|0\rangle|1\rangle)\ .\label{eq:W state}\end{eqnarray}
In this case the qubit values correspond to the occupation number
($0$ or $1$) of a spatial mode}, as described in Sections IIIA and
IIIB. The cluster state of \textcolor{black}{(\ref{eq:cluster state})}
is a superposition of two terms which correspond to GHZ states of
type \eqref{eq:GHZ} with\textcolor{black}{~$r=N/2$ and $r=N$}
respectively. From Fig. \ref{fig:Bell with r}, we note there is no
Bell violation when \textcolor{black}{$r=N$}. The effect is that
this cluster state case is less optimal than the symmetric GHZ state
($r=N/2$) and no violation of the $N=4$ CFRD inequality is observed.
\textcolor{black}{For W-states (\ref{eq:W state}), the value of the
$LHS$ of (\ref{eq:LHS}) is always $0$ so that no violations of
this particular CFRD inequality are found. Similarly negative results
were obtained for the arbitrary $N$ extensions of these superpositions
$|\psi\rangle=(|1\rangle^{\otimes N}+|0\rangle^{\otimes N/2}|1\rangle^{\otimes N/2}+|1\rangle^{\otimes N/2}|0\rangle^{\otimes N/2}-|0\rangle^{\otimes N})/2$
, where $N$ was increased.}

\textcolor{black}{We can also consider the specific four-photon cluster
state that has been produced experimentally by P. Walther et al \cite{clusterexpfour-photon}:}

\textcolor{black}{\begin{eqnarray}
|C\rangle & = & \frac{1}{2}(|H\rangle|H\rangle|H\rangle|H\rangle+|H\rangle|H\rangle|V\rangle|V\rangle\nonumber \\
 &  & +|V\rangle|V\rangle|H\rangle|H\rangle-|V\rangle|V\rangle|V\rangle|V\rangle)\ ,\label{eq:clusterHV}\end{eqnarray}
where $|H\rangle$ and $|V\rangle$ represent respectively horizontally
and vertically polarized single-photon states. }Thus, $|H\rangle\equiv|1\rangle_{H}|0\rangle_{V}$
and $|V\rangle\equiv|0\rangle_{H}|1\rangle_{V}$, the $|0\rangle_{H/V}$
and $|1\rangle_{H/V}$ being number states for each polarized mode
denoted by subscript $H$ and $V$. \textcolor{black}{The state (\ref{eq:clusterHVN})
is a four photon state and the qubit value is encoded in the polarisation.
We also consider the four-photon W-state:}

\textcolor{black}{\begin{eqnarray}
|W\rangle & = & \frac{1}{2}(|V\rangle|H\rangle|H\rangle|H\rangle+|H\rangle|V\rangle|H\rangle|H\rangle\nonumber \\
 &  & +|H\rangle|H\rangle|V\rangle|H\rangle+|H\rangle|H\rangle|H\rangle|V\rangle)\ .\label{eq:W stateHV}\end{eqnarray}
}As explained in Section 3A, these states can be converted to the
states $|\frac{N}{2}\rangle$ of (\ref{eq:N/2 state}) with $N=8$
via the apparatus of Fig. 1. We can then use the N=8 CFRD inequality
for testing the nonlocal behaviour of these states, which are written
explicitly as 

\textcolor{black}{\begin{eqnarray}
|C\rangle & = & \frac{1}{2}(|1\rangle|0\rangle|1\rangle|0\rangle|1\rangle|0\rangle|1\rangle|0\rangle+|1\rangle|0\rangle|1\rangle|0\rangle|0\rangle|1\rangle|0\rangle|1\rangle\nonumber \\
 &  & +|0\rangle|1\rangle|0\rangle|1\rangle|1\rangle|0\rangle|1\rangle|0\rangle\nonumber \\
 &  & -|0\rangle|1\rangle|0\rangle|1\rangle|0\rangle|1\rangle|0\rangle|1\rangle)\label{eq:clusterHVphoton}\end{eqnarray}
}and

\textcolor{black}{\begin{eqnarray}
|W\rangle & = & \frac{1}{2}(|0\rangle|1\rangle|1\rangle|0\rangle|1\rangle|0\rangle|1\rangle|0\rangle+|1\rangle|0\rangle|0\rangle|1\rangle|1\rangle|0\rangle|1\rangle|0\rangle\nonumber \\
 &  & \ \ +|1\rangle|0\rangle|1\rangle|0\rangle|0\rangle|1\rangle|1\rangle|0\rangle\nonumber \\
 &  & \ \ +|1\rangle|0\rangle|1\rangle|0\rangle|1\rangle|0\rangle|0\rangle|1\rangle)\ .\label{eq:WHVphoton-1}\end{eqnarray}
} \textcolor{black}{Evaluation of the left- and right- hand sides
of the CFRD Bell inequality (\ref{eq:CFRD_QM}) gives: $LHS=1/16$
and $RHS=3^{2}/2^{8}$. Thus, no violations are possible for the $N=8$
CFRD inequality with these 4 photon states so that in this case the
cluster state is less optimal than the $r=N/2$ GHZ state (\ref{eq:GHZ}). However
if we consider the $N$ photon extension of the 4-photon states (\ref{eq:clusterHVN})
and (\ref{eq:clusterHVphoton}):}

\textcolor{black}{\begin{eqnarray}
|C\rangle & = & \frac{1}{2}(|H\rangle^{\otimes N}+|H\rangle^{\otimes N/2}|V\rangle^{\otimes N/2}\nonumber \\
 &  & +|V\rangle^{\otimes N/2}|H\rangle^{\otimes N/2}-|V\rangle^{\otimes N})\ ,\label{eq:clusterHVN}\end{eqnarray}
 we find a more positive result: $LHS=1/16$ and $RHS=(3/4)^{N/2}$,
so that a violation is possible when $N\geq20$. We note in this case
no violations are obtained for the $W$ states. }

\subsection{Angular dependence}

The CFRD inequality \textcolor{black}{holds for arbitrary measurements
and is therefore not restricted to the orthogonal choice specified
in the previous section. Noting that Bell inequalities for correlated
spins are optimized at polarizer angles that are distinct at different
sites, we wish to investigate the effect of an arbitrary local oscillator
phase.}

\textcolor{black}{Using general quadrature operators $\hat{X}_{n}^{\theta}=(\hat{a}_{n}e^{-i\theta_{n}}+\hat{a}_{n}^{\dagger}e^{i\theta_{n}})/2$
for each site $n$, and defining $\hat{F}_{n}=\hat{X}_{n}^{\theta_{n}}+i\hat{X}_{n}^{\theta_{n}'}$,
we substitute these in the CFRD inequality \eqref{eq:CFRD} with $x_{n}=\hat{X}_{n}^{\theta}$
and $p_{n}=\hat{X}_{n}^{\theta'}$. The inequality then becomes}\textbf{\textcolor{black}{{}
}}\textcolor{black}{\begin{multline}
|\langle\prod_{n=1}^{N}F_{n}\rangle_{P}|^{2}\leq\left\langle \prod_{n=1}^{N}\left[a_{n}^{\dagger}a_{n}\right.\right.\\
\left.\left.+\frac{1}{4}a_{n}^{2}(e^{-2i\theta_{n}}+e^{-2i\theta_{n}'})+\frac{1}{4}a_{n}^{\dagger2}(e^{2i\theta_{n}}+e^{2i\theta_{n}'})\right]\right\rangle _{P},\end{multline}
which gives additional terms in the $RHS$. We note that in QM, these
extra terms correspond to operators that change the boson number by
two. Hence for states like $|0011\rangle+|1100\rangle$ they will
be zero, leaving the $RHS$ invariant with }the choice of phases $\theta_{n}$,
$\theta_{n}'$. Therefore, for GHZ states, the RHS will be just as
in Eq (\ref{eq:RHSvalue}).

For simplicity, we treat the case of $N=2$ as an illustration. Within
QM, the $LHS$ for the Bell state $|\psi\rangle=c_{1}|0\rangle|1\rangle+c_{2}|1\rangle|0\rangle$
is:\begin{eqnarray}
LHS & = & |\langle(\hat{X}_{1}^{\theta}+i\hat{X}_{1}^{\theta'})(\hat{X}_{2}^{\phi}+i\hat{X}_{2}^{\phi'})|^{2}\nonumber \\
 & = & |\langle[(\hat{x}_{1}\cos\theta+\hat{p}_{1}\sin\theta)+i(\hat{x}_{1}\cos\theta'+\hat{p}_{1}\sin\theta')]\nonumber \\
 & \times & [(\hat{x}_{2}\cos\phi+\hat{p}_{2}\sin\phi)+i(\hat{x}_{2}\cos\phi'+\hat{p}_{2}\sin\phi')]\rangle|^{2}\nonumber \\
 & = & |\frac{1}{2}c_{1}c_{2}[\left(cos(\theta-\phi)-cos(\theta'-\phi')\right)\nonumber \\
 &  & +i\left(cos(\theta'-\phi)+cos(\theta-\phi')\right)]|^{2}\ ,\end{eqnarray}
since here $\langle\hat{x}_{1}\hat{x}_{2}\rangle=\langle\hat{p}_{1}\hat{p}_{2}\rangle=c_{1}c_{2}/2$
and $\langle\hat{x}_{1}\hat{p}_{2}\rangle=\langle\hat{p}_{2}\hat{x}_{1}\rangle=0$.
The $LHS$ is maximised either with the orthogonal choice $\theta=\phi=0,\ \theta'=-\pi/2=-\phi',$
for which $\langle\mathrm{Im}\{\hat{F}_{1}\hat{F}_{2}\}\rangle=0$,
$\langle\mathrm{Re}\{\hat{F}_{1}\hat{F}_{2}\}\rangle=c_{1}c_{2}$\textcolor{black}{;
or with the choice $\theta=0,$ $\theta'=\pi/2$, $\phi=\pi/4$, $\phi'=-\pi/4$
for which $\langle\mathrm{Im}\{\hat{F}_{1}\hat{F}_{2}\}\rangle=c_{1}c_{2}/\sqrt{2}$,
$\langle\mathrm{Re}\{\hat{F}_{1}\hat{F}_{2}\}\rangle=c_{1}c_{2}/\sqrt{2}$.
In both cases, the $LHS$ is equal to $|c_{1}c_{2}|^{2}$, as in Eq
(\ref{eq:LHS}). Hence, just as in the orthogonal quadrature case,
there is no violation. }

The case for the extreme photon-number correlated state $|\psi\rangle=c_{1}|0\rangle|0\rangle+c_{2}|1\rangle|1\rangle$
is even less favourable. The value of $LHS$ can still be \textcolor{black}{maximized
to $|c_{1}c_{2}|^{2}$, with angle choice $\theta=\phi=0,$ $\theta'=\phi'=\pm\pi/2$
or $\theta=0$, $\theta'=\pi/2$, $\phi=-\pi/4$, $\phi'=\pi/4$.
However, the value of the $RHS$ is increased even further, so that
again, there is no violation. }

\textcolor{black}{More generally, the inequality is violated for $N\ge10$,
that is, with more correlated modes and observers. We find that the
optimal violations are found with phase choices: $\theta_{1}=\cdots=\theta_{n}=0,$
and $\theta_{1}^{'}=\cdots=\theta_{r}^{'}=-\theta_{r+1}^{'}=\cdots=-\theta_{n}^{'}=-\pi/2$
or $\theta_{n}=(-1)^{n+1}\pi(n-1)/(2N)$, $\theta_{n}^{'}=\theta_{n}+\pi/2\ (n\leq r)$
and $\theta_{n}=(-1)^{n}\pi(n-1)/(2N)$, $\theta_{n}^{'}=\theta_{n}-\pi/2\ (n>r)$.
The $LHS$ and $RHS$ then become, for arbitrary $N$ and $r$, with
$c_{1}=c_{2}=1/\sqrt{2}$: \begin{eqnarray}
{\color{black}LHS_{max}} & {\color{black}=} & {\color{black}\mid c_{1}c_{2}\mid^{2}=1/4\ ,}\nonumber \\
{\color{black}RHS}_{max} & {\color{black}=} & {\color{black}\frac{3^{N-r}+3^{r}}{2^{N+1}}\ .}\end{eqnarray}
}

In summary, we find that \textcolor{black}{the violations are }\textcolor{black}{\emph{independent}}\textcolor{black}{{}
of the relative phases between different sites} \cite{functional letter}---unlike
the usual case with Bell inequalities for correlated spins---, thus
confirming a result pointed out in \cite{cvbell2} for this larger
range of states.

\section{MABK inequalities}

In early proposals for obtaining inequalities to test local realism
with continuous variable measurements, only the sign of the quadrature
variable is recorded \cite{cvbin}. As far as LHV theories are concerned,
it is just the final outcomes that are relevant, not the quantum operators
corresponding to them. This allows one to use the same logic that
is employed for dichotomic spin variables. Our results here are therefore
similar to those of earlier workers \cite{cvbin,munro,Acin }. As
in the work of \cite{Acin }, we find an exponential improvement in
Bell violation with increasing number of sites, but we demonstrate
that result for a larger class of states, some of which may be within
experimental feasibility.

\subsection{MABK approach with binned variables}

Violations of the MABK inequalities, which were originally used for
spin variables, can also be obtained by using binned outcomes of continuous
variable observables. These are applicable to any quantum state. Here
the binned measurement $x_{n}^{bin}=2\Theta\left(x_{n}\right)-1$
is the observable with outcome $x_{n}^{bin}=+1$, if $x_{n}\geq0$,
and $x_{n}^{bin}=-1$, if $x_{n}<0$. We define $p_{n}^{bin}$ similarly,
together with\textcolor{black}{{} corresponding quantum operators }$\hat{x}_{n}^{bin}=2\Theta\left(\hat{x}_{n}\right)-1$\textcolor{black}{{}
and }$\hat{p}_{n}^{bin}=2\Theta\left(\hat{p}_{n}\right)-1$. Next,
we can introduce a new binned complex observable $a_{n}^{bin}$ at
location $n$ through its real and imaginary parts, so that $\mathrm{Re\{}a_{n}^{bin}\}\equiv x_{n}^{bin}$
and $\mathrm{Im}\{a_{n}^{bin}\}=p_{n}^{bin}$. There is also a corresponding
operator, $\hat{a}_{n}^{bin}=\hat{x}_{n}^{bin}+i\hat{p}_{n}^{bin}$.

For $N$ sites, we can modify the CFRD proof directly, assuming local
hidden variables, to get:

\begin{equation}
\left|\left\langle a_{1}^{bin}a_{2}^{bin}...\right\rangle _{P}\right|^{2}\leq\left\langle \prod_{n=1}^{N}(\left[x_{n}^{bin}\right]^{2}+\left[p_{n}^{bin}\right]^{2})\right\rangle _{P}=2^{N}.\end{equation}
This follows because $\langle\left[x_{n}^{bin}\right]^{2}\rangle_{P},\langle\left[p_{n}^{bin}\right]^{2}\rangle_{P}=1$.
The inequality remains valid if any of the complex variables $a_{n}^{bin}$
are replaced by their conjugates. 

\textcolor{black}{The CFRD inequality for this binned case then reduces
to that of Mermin and Ardehali \cite{mermin} by taking the square
root of both of sides. However, the inequality was later strengthened
by Belinskii and Klyshko\cite{Klyshko}, to give an approach which
we refer to as the MABK  inequality. This approach can be extended
to account for more general angles and states, by introducing binned
angular observables }$X_{n}^{\theta,bin}=2\Theta\left(X_{n}^{\theta}\right)-1$.\textcolor{black}{{}
Next, $a_{n}^{bin}$ is replaced with $F_{n}^{bin}=X_{n}^{\theta,bin}+iX_{n}^{\theta',bin}$.
The notation $\theta_{n}$ and $\theta'_{n}$ represents the angles
defining the two local oscillator phases chosen at site $n$, where
$k=1,...,N$, giving $2^{N}$ distinct sets of measurement settings. }

\textcolor{black}{Then, defining $\Pi_{N}=F_{1}^{bin}F_{2}^{bin}...F_{N}^{bin}$,
we can test for violations of the even stronger MABK inequality \begin{equation}
\left|S_{N}\right|\leq1\ ,\label{eq:MK}\end{equation}
where \begin{equation}
S_{N}=2^{-N/2}\left\langle \mathrm{Re}\{\Pi_{N}\}\pm\mathrm{Im\{}\Pi_{N}\}\right\rangle _{P}\end{equation}
, for $N$ even, and \begin{equation}
S_{N}=2^{-(N-1)/2}\mathrm{Re}(\mathrm{Im})\left\{ \left\langle \Pi_{N}\right\rangle _{P}\right\} \end{equation}
for $N$ odd. The inequality also holds for $S_{N}=2^{-(N-1)/2}\left\langle \sqrt{(\mathrm{Re}\{\Pi_{N}\})^{2}+(\mathrm{Im}\{\Pi_{N}\})^{2}}\right\rangle _{P}$.
These inequalities have been considered recently for the case of GHZ
states (\ref{eq:state}), where $r=N$, by Acin }\textcolor{black}{\emph{et
al}}\textcolor{black}{{} \cite{Acin }. We can also define $\Pi_{N}$
with the adjoint operators at any of the sites, to obtain the same
inequality. }

\subsection{Three-site example}

\textcolor{black}{Within quantum mechanics, the $LHS$ of the Bell
inequality (\ref{eq:MK}) can be calculated via binned quadrature
phase amplitudes. It is instructive to give an example of the calculation
in detail. For the state (\ref{eq:state}), the lowest values of $N$
and $r$ that give a violation are $N=3$ and $r=1$: \begin{equation}
|\psi\rangle=\frac{1}{\sqrt{2}}(|0\rangle_{1}|1\rangle_{2}|1\rangle_{3}+|1\rangle_{1}|0\rangle_{2}|0\rangle_{3})\ .\end{equation}
Using the harmonic oscillator wavefunctions given in the Appendix,
we find that:} \begin{eqnarray}
\langle\hat{\Pi}_{3}\rangle & = & \langle\hat{F}_{1}^{bin}\hat{F}_{2}^{bin}\hat{F}_{3}^{bin}\rangle\nonumber \\
 & = & (\frac{2}{\pi})^{3/2}[cos(\theta-\phi-\gamma)-cos(\theta-\phi'-\gamma')\nonumber \\
 &  & -cos(\theta'-\phi'-\gamma)-cos(\theta'-\phi-\gamma')\nonumber \\
 &  & +i(cos(\theta-\phi'-\gamma)+cos(\theta-\phi-\gamma')\nonumber \\
 &  & +cos(\theta'-\phi-\gamma)-cos(\theta'-\phi'-\gamma'))]\ ,\label{eq:I_3}\end{eqnarray}
\textcolor{black}{where we denote $\theta_{1}=\theta$, $\theta_{2}=\phi$,
and $\theta_{3}=\gamma$. }

\textcolor{black}{Choosing the two measurements made at each site
to be orthogonal, and measurements at all sites along the same (or
opposite) directions, so that}\textcolor{green}{{} }$\ \theta=\phi=\gamma=0$,
and $\theta'=-\phi'=-\gamma'=-\pi/2\ $ (which means that $\Pi_{3}=a_{bin}^{\dagger}b_{bin}c_{bin}$),
we obtain

\textcolor{black}{\begin{equation}
S_{3}=2^{-(3-1)/2}\sqrt{(\mathrm{Re}\{\Pi_{3}\})^{2}+0}=2(2/\pi)^{3/2}>1\ .\label{eq:anlgeorth}\end{equation}
Alternatively, using different measurement angles: $\theta=0,\ \phi=-\pi/6,\ \gamma=-2\pi/6,\ \theta'=\pi/2,\ \phi'=-4\pi/6,\ \gamma'=-5\pi/6,\ $we
can rotate the $LHS$ moment, so that \begin{equation}
S_{3}=2^{-(3-1)/2}\sqrt{0+(\mathrm{Im}\{\Pi_{3}\})^{2}}=2(2/\pi)^{3/2}>1\ .\label{eq:anglenonorth}\end{equation}
}

\textcolor{black}{Next, consider the extreme GHZ state of form (\ref{eq:state})
with $r=3$:}

\textcolor{black}{\begin{equation}
|\psi\rangle=\frac{1}{\sqrt{2}}(|0\rangle_{1}|0\rangle_{2}|0\rangle_{3}+|1\rangle_{1}|1\rangle_{2}|1\rangle_{3})\ .\end{equation}
We obtain a new $\langle\Pi_{3}\rangle$ with $cos(\theta+\phi+\gamma)$,
$cos(\theta+\phi'+\gamma')$, etc.~replacing $cos(\theta-\phi-\gamma)$,
$cos(\theta-\phi'-\gamma')$, etc.~in Eq.~(\ref{eq:I_3}). In this
case the choice $\theta=0,\ \phi=\pi/6,\ \gamma=2\pi/6,\ \theta'=\pi/2,\ \phi'=4\pi/6,\ \gamma'=5\pi/6,\ $
gives the maximum violation of the inequality,}

\textcolor{black}{\begin{equation}
S_{3}=2^{-(3-1)/2}\sqrt{(\mathrm{Re}\{\Pi_{3}\})^{2}+0}=2(2/\pi)^{3/2}>1\ ,\end{equation}
as has been presented by Acin }\textit{\textcolor{black}{et al.}}

\subsection{N-site inequalities}

Continuing in this fashion for higher $N$, we find that \begin{equation}
S_{N}=\frac{\sqrt{2}}{2}(\frac{4}{\pi})^{N/2}\end{equation}
for th\textcolor{black}{e choice}\textcolor{red}{{} $\ {\color{black}\theta_{n}=(-1)^{N+1}\pi(k-1)/(2N)}$}\textcolor{black}{,
}$\theta_{n}^{'}=\theta_{n}+\pi/2$ for $k\leq r$, and $\theta_{n}=(-1)^{N}\pi(n-1)/(2N)$,
$\theta_{n}^{'}=\theta_{n}-\pi/2$ for $n>r$. There is a violation
of the Bell inequality when $S_{N}>1$\textcolor{black}{.} This result
has been presented by Acin \textit{\textcolor{black}{et al.}} for
the case of $r=N$. We confirm the exponential increase with number
of sites $N$, but also make the observation that the violation occurs
for all types of states of the form (\ref{eq:state}), independently
of $r$. This contrasts with the result for the CFRD inequality, which
is strongly dependent on $r$, requiring $r\sim N/2$. Violation of
the BI with binning is therefore possible in principle for $N\geq3$,
but, as shown in later sections, this strategy is very sensitive to
losses and noise.

\subsection{Cluster and W states}

\textcolor{black}{We also present initial results for the violations
of the MABK CV inequalities using the specific cluster and W-states
(\ref{eq:cluster state}) and (\ref{eq:W state}) We first consider
the case $N=4$ with CV binned outcomes. Using the same method as
above, and selecting the optimal choice of angles, we obtain the best
N=4 MABK correlations of\begin{eqnarray}
|S_{4-C}| & = & \frac{\sqrt{2}}{4}\left(\frac{4}{\pi}\right)^{2}<|S_{4-GHZ}|=\frac{\sqrt{2}}{2}\left(\frac{4}{\pi}\right)^{2}\ ,\nonumber \\
|S_{4-W}| & = & 0<|S_{4-GHZ}|\ ,\end{eqnarray}
for cluster and W-states respectively, and thus we see no violation
here for $N=4$. However, on examining the generalisation \begin{equation}
|\psi\rangle=(|1\rangle^{\otimes N}+|0\rangle^{\otimes N/2}|1\rangle^{\otimes N/2}+|1\rangle^{\otimes N/2}|0\rangle^{\otimes N/2}-|0\rangle^{\otimes N/2})/2\label{eq:supclusterext}\end{equation}
of the superpositions (\ref{eq:cluster state}) and (\ref{eq:W state})
for higher $N$, we find $|S_{N-C}|=\frac{\sqrt{2}}{4}\left(\frac{4}{\pi}\right)^{N/2}$
so that violations are possible for $N\geq10$. Calculation for the
W-states shows no violation for any $N$. These calculations are specific
to the case where the qubit is realized as a photon number of $1$
or $0$ at a given spatial mode. The cases (\ref{eq:clusterHVN})
- (\ref{eq:WHVphoton-1}) where the qubit is realized in terms of
horizontal or vertical polarisation, as explained in Section III,
could also be examined.We leave open the question of whether other
Bell inequalities might be more sensitive for the cluster and W-states
with continuous variables measurements, as shown by Scarani et al
\cite{clusternonlocal} for spin measurements.}

\section{Functional optimization}

In this section, we show that greatly improved experimental simplicity,
together with robustness against decoherence is possible by testing
local realism using a\emph{ functional} moment inequality approach.
T\textcolor{black}{he advantage of the CFRD moment approach compared
to the binning approach \cite{cvbin} is that it gives a low weight
to quadrature signals with low amplitude, which reduces the sensitivity
to noise. On the other hand, binning has the advantage that it saturates,
which increases the relative size of the violations. We show that
}an approach that combines these two features using \emph{functions}
of quadrature variables is possible. 

\textcolor{black}{We analytically calculate the optimal function }using
a variational calculus\textcolor{black}{{} method. This produces an
inequality which is violated by the states of Eq. (\ref{eq:GHZ})
for $N\geq5$, i.e. half the number of modes required with the method
of \cite{cvbell2}. The violation increases exponentially with $N$,
and we will see that the critical detection efficiency $\eta_{crit}$
decreases asymptotically to $0.69$, a significant reduction.}

\subsection{The functional inequalities}

Functional inequalities are already implicit in the original derivation
of CFRD \textcolor{black}{\cite{cvbell2}}. For completeness, we present
a proof that takes explicit account of functions of measurements that
can be made at each of $N$ causally separated sites. We consider
the measurable joint probability $P(X_{1}^{\theta},X_{2}^{\phi}...)$
for outcome $X_{1}^{\theta}$, $X_{2}^{\phi}$, $\ldots$ at locations
$1$, $2$, $\ldots$, respectively, where $\theta$, $\phi$, $\ldots$
represents a choice of measurement parameter. For local hidden variable
(LHV) theories, the joint probability is written \cite{Bell} in terms
of variables $\lambda$, as \begin{equation}
P(X_{1}^{\theta},X_{2}^{\phi},...)=\int_{\lambda}d\lambda P(\lambda)P(X_{1}^{\theta}|\lambda)P(X_{2}^{\phi}|\lambda)...\ ,\label{eqn:sepprob}\end{equation}
where $P(X_{1}^{\theta}|\lambda)$ is the probability for result $X_{1}^{\theta}$
given that the system is specified by the hidden variable state $\lambda$.
We note that while we use the term {}``hidden variable state'',
the states can in fact be quantum states, denoted $\hat{\rho}_{\lambda}$,
as long as they are separable: i.e., not entangled. The assumption
of local causality (or separability) allows the factorization in the
integrand. From this, it also follows that we can write the measurable
moments $\langle X_{1}X_{2}X_{3}...\rangle$, where $X_{n}$ is an
observable associated with a measurement at site $n$, etc., as \begin{equation}
\langle X_{1}^{\theta}X_{2}^{\phi}X_{3}^{\gamma}...\rangle_{P}=\int_{\lambda}d\lambda P(\lambda)\langle X_{1}^{\theta}\rangle_{\lambda}\langle X_{2}^{\phi}\rangle_{\lambda}\langle X_{3}^{\gamma}\rangle_{\lambda}...\ .\label{eq:LHV_correlations}\end{equation}
Here $\langle X_{n}^{\theta}\rangle_{\lambda}$ is the average of
$X_{n}^{\theta}$ given the local hidden variable state $\lambda$,
which means that \begin{equation}
\langle X_{n}^{\theta}\rangle_{\lambda}=\int dX_{n}^{\theta}P(X_{n}^{\theta}|\lambda)\, X_{n}^{\theta}\ ,\end{equation}
where $X_{n}^{\theta}$ denotes the outcome of measurement at site
$n$ with phase $\theta$. 

Next we construct, for each site $n$, real functions of the two observables
$f_{n}(X_{n}^{\theta})$, $g_{n}(X_{n}^{\theta'})$, and define the
complex function:\begin{equation}
F_{n}=f_{n}(X_{n}^{\theta})+ig_{n}(X_{n}^{\theta'})\ .\end{equation}
The complex moment $\langle F_{1}F_{2}...F_{N}\rangle$ can be expressed
using sums of real-valued expressions of the type $\langle f_{1}(X_{1}^{\theta})g_{2}(X_{2}^{\phi'})...f_{N}(X_{N}^{\varphi})\rangle$.
Of course, $f_{n}(X_{n}^{\theta})$ is just another observable composed
of the observable $X_{n}^{\theta}$ plus some local post-measurement
processing. This post-processing can be carried out at any time subsequent
to the measurement. Eq. \eqref{eq:LHV_correlations} is therefore
valid for \begin{equation}
\langle f(X_{1}^{\theta})\ldots f(X_{N}^{\varphi})\rangle_{P}=\int_{\lambda}d\lambda P(\lambda)\langle f(X_{1}^{\theta})\rangle_{\lambda}\ldots\langle f(X_{N}^{\varphi})\rangle_{\lambda}.\end{equation}
 The expectation value of products of the $F_{n}$ must satisfy:

\begin{equation}
\langle F_{1}\ldots F_{N}\rangle_{P}=\int_{\lambda}d\lambda P(\lambda)\langle F_{1}\rangle_{\lambda}\ldots\langle F_{N}\rangle_{\lambda}\ ,\label{eq:F_k_moments}\end{equation}
where $\langle F_{n}\rangle_{\lambda}\equiv\langle f_{n}(X_{n}^{\theta})\rangle_{\lambda}+i\langle g_{n}(X_{n}^{\theta'})\rangle_{\lambda}.$
Hence, from \eqref{eq:F_k_moments}, the following inequality holds:

\begin{eqnarray}
|\langle F_{1}F_{2}...F_{N}\rangle_{P}|^{2} & \leq & \int d\lambda P(\lambda)|\langle F_{1}\rangle_{\lambda}|^{2}...|\langle F_{N}\rangle_{\lambda}|^{2}\ .\label{eq:F_k_mod_squared}\end{eqnarray}
For any particular value of $\lambda,$ the statistics predicted for
$f_{n}(X_{n})$ must have a non-negative variance, i.e., $\langle f_{n}(X_{n})\rangle_{\lambda}^{2}\leq\langle f_{n}(X_{n})^{2}\rangle_{\lambda}$.
Then we can rewrite \eqref{eq:F_k_mod_squared} explicitly in terms
of the $f_{n}$'s. Using this variance inequality we arrive at a functional
moment inequality:\begin{equation}
B=\frac{\left|\left\langle \prod_{n=1}^{N}\left[f_{n}(X_{n}^{\theta})+ig_{n}(X_{n}^{\theta'})\right]\right\rangle \right|}{\left\langle \prod_{n=1}^{N}\left[f_{n}(X_{n}^{\theta})^{2}+g_{n}(X_{n}^{\theta'})^{2}\right]\right\rangle }^{2}\leq1\ .\label{DUPLICATE: eq:moment cv BI}\end{equation}
\textcolor{black}{We will measure the violation of this inequality
by the Bell observable $B$ defined above, so that failure of LHV
is demonstrated when $B>1$. }

\subsection{\textcolor{black}{Functional inequality examples}}

\begin{figure}[h]
\includegraphics[width=1\columnwidth]{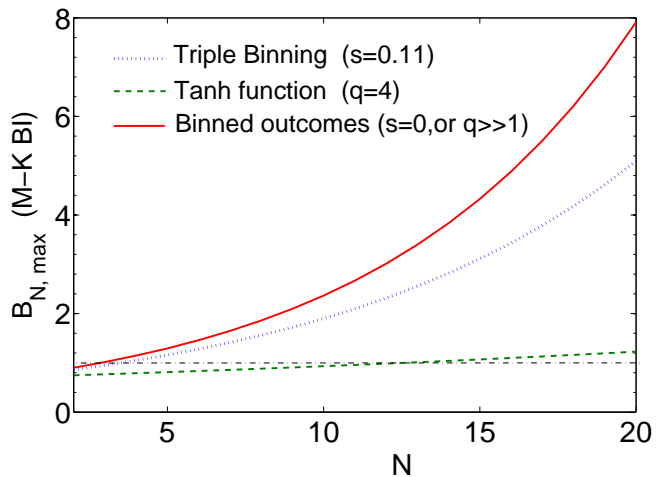}

\caption{\textcolor{black}{(Color online) Maximum violations of MABK Bell inequalities
as a function of the number of modes $N$ with ideal detection efficiency.
Violation of the Bell inequalities is achieved when $B>1$. }The triple
step function case is the dotted line with $s=0.11$, which reduces
to the binned case when $s=0$ (solid line); the $tanh$ function
case is the dashed line with $q=4$, $m=1.4$, which reduces to the
binned case when $q$ is large. Here the parameters $m$, $q$, and
$s$ are the optimal choices to maximize the Bell value.\label{fig:MK-ideal-1}}

\end{figure}

\textcolor{black}{First, we investigate three specific types of function: }
\begin{enumerate}
\item \textcolor{black}{Fractional order moments: $f_{n}(x)=g_{n}(x)=|x|^{m}sign(x)$.}
\item Triple binning\textcolor{black}{: $\ f_{n}(x)\ $ is the triple-valued
step function equal to $-1$ when $x<-s$, $+1$ when $x>s$, and
$0$ elsewhere.}
\item \textcolor{black}{Powers of }\textcolor{black}{\emph{$tanh$$\ $}}\textcolor{black}{functions:
$f_{n}(x)=g_{n}(x)=sign(x)|\tanh(qx)|^{m}$. }
\end{enumerate}
\textcolor{black}{For all of these functions, we obtain an LHV violation
when $N\geq5$, with optimal choice of $m$, $s$, and $q$. This
is a much more feasible experiment than that for an integer moment
with $m=1$, which required $N\geq10$ in CFRD. The reason for this
is that these three functions all combine high intensity saturation
with small-noise insensitivity. The region close to $0$ is relatively
flat, rather than discontinuous as in the binning approach. This leads
to reduced sensitivity to small noise effects, which otherwise will
lead to a greatly amplified contribution of random vacuum noise to
the correlations. Here, we again select the optimal $r$ at $r=N/2$
($N$ is even) or at $r=(N\pm1)/2$ ($N$ is odd). The extreme state
where $r=N$ is less optimal.}

\textcolor{black}{It is worth noting that bounded CV observables (cases
2 and 3 above) }can also be used with the MABK inequalities. These
also reduce to the binning results asymptotically$\ $as $s=0$ in
the triple binning case and as $q\rightarrow\infty$ or $m\rightarrow0$
in the \textit{tanh} case, as shown in Fig. \ref{fig:MK-ideal-1}.
Small $q$ is less optimal as shown by dashed curve. However, for
the MABK inequalities the triple binning case ($s\neq0$) is less
optimal than the standard binned case ($s=0$) even if we select a
nonzero optimal value of $s$. 

\textcolor{black}{}%
\begin{figure}[h]
\textcolor{black}{\includegraphics[width=1\columnwidth,height=6.2cm]{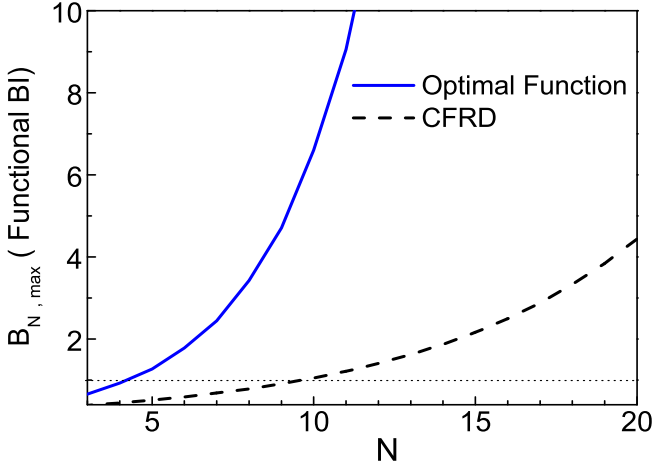}}

\textcolor{black}{\caption{\textcolor{black}{(Color online) Maximum violations of functional
CV inequality with GHZ states} as a function of the number of modes
$N$.~Violation of the Bell inequality is achieved when $B>1$.~The
violations using~\textcolor{black}{{} the optimal function (solid)}
are much stronger than the CFRD result (dashed). \label{fig:ideal Bell value-1}}
}
\end{figure}

\subsection{Optimized functional CFRD}

\textcolor{black}{In order to get the strongest violation of LHV theories,
we can optimize the function of observables by using variational calculus:
\begin{equation}
\frac{\delta B}{\delta f_{n}}=\frac{\delta B}{\delta g_{n}}=\frac{\delta}{\delta f_{n}}\left[\frac{|\langle\prod_{i=1}^{N}\bigl\{ f_{n}(\hat{X}_{n}^{\theta})+ig_{n}(\hat{X}_{n}^{\theta'})\bigl\}\rangle|^{2}}{\langle\prod_{i=1}^{N}\bigl\{ f_{n}(\hat{X}_{n}^{\theta})^{2}+g_{n}(\hat{X}_{n}^{\theta'})^{2}\bigl\}\rangle}\right]=0\ .\label{eq:optimal function}\end{equation}
For simplicity, we assume the functions $f_{n}$ and $g_{n}$ are
odd. The numerator can be maximized by choosing orthogonal angles,
while the denominator of the fraction is invariant with angles. In
the case of the GHZ states, we find that:}

\begin{equation}
B_{N}=\frac{2^{N-1}(\frac{2}{\pi})^{\frac{N}{2}}(\prod_{n=1}^{N}I_{n}^{+}+\prod_{n=1}^{N}I_{n}^{-})^{2}}{\prod_{n=1}^{r}I_{n}\prod_{k=r+1}^{N}I_{n}^{o}+\prod_{n=1}^{r}I_{n}^{o}\prod_{k=r+1}^{N}I_{n}}\ ,\label{DUPLICATE: DUPLICATE: eq:ideal symmetric}\end{equation}
\textcolor{black}{where $I_{n}^{\pm}=2\int e^{-2x^{2}}xf_{n}^{\pm}dx$,
$I_{n}=4\int x^{2}e^{-2x^{2}}[(f_{n}^{+})^{2}+(f_{n}^{-})^{2})]dx$,
and $I_{n}^{o}=\int e^{-2x^{2}}[(f_{n}^{+})^{2}+(f_{n}^{-})^{2})]dx$
(given in the Appendix) are different integrals for $x$ which contribute
to the expectation values in both sides of inequality (\ref{DUPLICATE: eq:moment cv BI}).
Here $f_{n}^{\pm}=f_{n}\pm g_{n}$, and the factor }$e^{-2x^{2}}$\textcolor{black}{{}
~was obtained from the joint probability of observables. Requiring
$\delta B_{N}/\delta f_{n}^{\pm}=0$$ $, we find the optimal condition:
$f_{n}(x)=\pm g_{n}(x)$. The components of complex functions $f_{n}$,
$g_{n}$ are the same at each site, and have the form\begin{equation}
f_{n}(x)=g_{n}(x)=\frac{x}{1+\varepsilon_{N}x^{2}}\ .\label{eq:optimum_function}\end{equation}
For the even $N$ case, it is optimal to choose $r=N/2$. Then $\varepsilon_{N}$
is independent of $N$, but has to be calculated numerically since
it satisfies a nonlinear integral equation: $\varepsilon_{N}=4I^{o}/I$.
Using the method shown in the Appendix, we can obtain the fixed values
of integrals. We find that $\varepsilon_{N}=2.9648$ gives the optimal
case, and $B_{N}=2^{N-2}\left[2\left(I^{+}\right)^{4}/(\pi I^{o}I)\right]^{N/2}$. }

\textcolor{black}{For $N$ an odd number, }the greatest violations
occur for $r=(N-1)/2$.~\textcolor{black}{{} The optimal function has
the same form as in (\ref{eq:optimum_function}) except that the parameter
$\varepsilon_{N}$ changes to $\varepsilon'_{N}$, where: \begin{equation}
\varepsilon'{}_{N}\equiv\varepsilon_{N}[\frac{N\varepsilon_{N}^{+}-\varepsilon_{N}^{-}}{N\varepsilon_{N}^{+}+\varepsilon_{N}^{-}}]\ ,\label{eq:odd_optimize}\end{equation}
and $\varepsilon_{N}^{\pm}=\varepsilon_{N}\pm4$. However, the numerical
value of $\varepsilon_{N}$ and $\varepsilon'_{N}$ now depend on
$N$, as the integral equation (\ref{eq:odd_optimize}) for odd values
of $N$ is $N$-dependent. This provides better violations of \eqref{DUPLICATE: eq:moment cv BI}
than any other arbitrary function, provided $N\geq5$. The maximum
$B_{N}$ value with this optimal choice is shown in Fig. \ref{fig:ideal Bell value-1},
compared with the CFRD result which uses a simple correlation function. }

In summary, a functional optimization approach gives a substantially
larger violation of the CFRD inequality, and we have carried this
optimization out explicitly for the case of a generalized GHZ state.

\section{Many observables per site: SV inequalities}

\textcolor{black}{In this Section, we compare the CFRD inequality
with the inequalities given by Shchukin and Vogel (SV) \cite{Shchukin}.
These generalize the CFRD inequality to measurements with $4$ or
$8$ settings or observables at each site. In principle, this may
allow larger violations or greater robustness against loss and noise.
However, these are increasingly more complicated both to analyse and
to carry out experiments for. By comparison, the CFRD approach considers
just two detector settings per site.}

\subsection{Four setting SV inequality}

\textcolor{black}{First, we evaluate the results of SV for $4$ observables
per site, using arbitrary choices of phase for the quadrature measurements,
where $F_{n}=(X_{n}^{\theta_{n}^{1}})^{m}+i(X_{n}^{\theta_{n}^{2}})^{m}+j(X_{n}^{\theta_{n}^{3}})^{m}+k(X_{n}^{\theta_{n}^{4}})^{m}$,
are quaternionic functions of observables. Here $X_{n}^{\theta_{n}^{\ell}}$,
$(\ell=1,\ 2,\ 3,\ 4)$ are real observables at site $n$. The multiplication
rules for the quaternionic units $i,\ j$, and $k$ are $i^{2}=j^{2}=k^{2}=-1,$
$ij=-ji=k,\ jk=-kj=i,\ ki=-ik=j$. Thus, Shchukin and Vogel (SV) \cite{Shchukin}
obtain the following inequality within any hidden-variable theory:
\begin{eqnarray}
|\langle F_{1}F_{2}...\rangle_{P}|^{2} & \leq & \langle\sum_{l=1}^{4}(X_{1}^{\theta_{1}^{\ell}})^{2m}\times\sum_{l=1}^{4}(X_{2}^{\theta_{2}^{\ell}})^{2m}...\rangle_{P}\ .\end{eqnarray}
We consider the class of entangled} states of type (\ref{eq:GHZ})
to \textcolor{black}{find violations of this inequality. }We use the
optimal parameters found for the CFRD inequality, $r=N/2$ and $c_{1}=c_{2}=1/\sqrt{2}$,
for comparison with the SV inequalities with $4$ observables per
site.

\subsubsection{Two-site case}

\textcolor{black}{As these inequalities are more complicated theoretically,
we begin with the simplest example of $N=2$. At site $n=1$, we have
$4$ choices of observables $(X_{A}^{\theta_{1}^{l}})^{m}$ ($l=1,\ 2,\ 3,\ 4$)
with corresponding outcomes $|x_{A}^{\theta_{1}^{l}}|^{m}sign(x_{A}^{\theta_{1}^{l}})$,
and similarly for the other site. The values of $LHS$ and $RHS$
are evaluated in terms of integrals for $x$. }First we evaluate the
inequalities for $N=2$, with the state $|\psi\rangle=c_{1}|0\rangle|1\rangle+c_{2}|1\rangle|0\rangle$,
using \textcolor{black}{$c_{1}=c_{2}=1/\sqrt{2}$ and $m=1/3$. The
LHS of the SV inequality has many terms:}

\textcolor{black}{\begin{eqnarray}
LHS & = & |\langle[(\hat{X}_{A}^{\theta^{1}})^{\frac{1}{3}}+i(\hat{X}_{A}^{\theta^{2}})^{\frac{1}{3}}+j(\hat{X}_{A}^{\theta^{3}})^{\frac{1}{3}}+k(\hat{X}_{A}^{\theta^{4}})^{\frac{1}{3}}]\times\nonumber \\
 &  & \times[(\hat{X}_{_{B}}^{\phi^{1}})^{\frac{1}{3}}+i(\hat{X}_{_{B}}^{\phi^{2}})^{\frac{1}{3}}+j(\hat{X}_{_{B}}^{\phi^{3}})^{\frac{1}{3}}+k(\hat{X}_{_{B}}^{\phi^{4}})^{\frac{1}{3}}]\rangle|^{2}\nonumber \\
 & = & \mid\langle[(\hat{X}_{A}^{\theta^{1}})^{\frac{1}{3}}(\hat{X}_{B}^{\phi^{1}})^{\frac{1}{3}}-(\hat{X}_{A}^{\theta^{2}})^{\frac{1}{3}}(\hat{X}_{B}^{\phi^{2}})^{\frac{1}{3}}\nonumber \\
 &  & -(\hat{X}_{A}^{\theta^{3}})^{\frac{1}{3}}(\hat{X}_{B}^{\phi^{3}})^{\frac{1}{3}}-(\hat{X}_{A}^{\theta^{4}})^{\frac{1}{3}}(\hat{X}_{B}^{\phi^{4}})^{\frac{1}{3}}]\nonumber \\
 &  & +i[(\hat{X}_{A}^{\theta^{2}})^{\frac{1}{3}}(\hat{X}_{B}^{\phi^{1}})^{\frac{1}{3}}+(\hat{X}_{A}^{\theta^{1}})^{\frac{1}{3}}(\hat{X}_{B}^{\phi^{2}})^{\frac{1}{3}}\nonumber \\
 &  & -(\hat{X}_{A}^{\theta^{4}})^{\frac{1}{3}}(\hat{X}_{B}^{\phi^{3}})^{\frac{1}{3}}+(\hat{X}_{A}^{\theta^{3}})^{\frac{1}{3}}(\hat{X}_{B}^{\phi^{4}})^{\frac{1}{3}}]\nonumber \\
 &  & +j[(\hat{X}_{A}^{\theta^{3}})^{\frac{1}{3}}(\hat{X}_{B}^{\phi^{1}})^{\frac{1}{3}}+(\hat{X}_{A}^{\theta^{4}})^{\frac{1}{3}}(\hat{X}_{B}^{\phi^{2}})^{\frac{1}{3}}\nonumber \\
 &  & +(\hat{X}_{A}^{\theta^{1}})^{\frac{1}{3}}(\hat{X}_{B}^{\mathbf{\phi^{3}}})^{\frac{1}{3}}-(\hat{X}_{A}^{\theta^{2}})^{\frac{1}{3}}(\hat{X}_{B}^{\phi^{4}})^{\frac{1}{3}}]\nonumber \\
 &  & +k[(\hat{X}_{A}^{\theta^{4}})^{\frac{1}{3}}(\hat{X}_{B}^{\phi^{1}})^{\frac{1}{3}}-(\hat{X}_{A}^{\theta^{3}})^{\frac{1}{3}}(\hat{X}_{B}^{\phi^{2}})^{\frac{1}{3}}\nonumber \\
 &  & +(\hat{X}_{A}^{\theta^{2}})^{\frac{1}{3}}(\hat{X}_{B}^{\phi^{3}})^{\frac{1}{3}}+(\hat{X}_{A}^{\theta^{1}})^{\frac{1}{3}}(\hat{X}_{B}^{\phi^{4}})^{\frac{1}{3}}]\rangle\mid^{2}\ ,\end{eqnarray}
}

\textcolor{black}{Similarly, the RHS is:\begin{eqnarray}
RHS & = & \langle\bigl\{(\hat{X}_{A}^{\theta^{1}})^{2/3}+(\hat{X}_{A}^{\theta^{2}})^{2/3}+(\hat{X}_{A}^{\theta^{3}})^{2/3}+(\hat{X}_{A}^{\theta^{4}})^{2/3}\bigl\}\nonumber \\
 &  & \bigl\{(\hat{X}_{B}^{\phi^{1}})^{2/3}+(\hat{X}_{B}^{\phi^{2}})^{2/3}+(\hat{X}_{B}^{\phi^{3}})^{2/3}+(\hat{X}_{B}^{\phi^{4}})^{2/3}\bigl\}\rangle\nonumber \\
 & = & \langle(\hat{X}_{A}^{\theta^{1}})^{2/3}(\hat{X}_{B}^{\phi^{1}})^{2/3}+(\hat{X}_{A}^{\theta^{1}})^{2/3}(\hat{X}_{B}^{\phi^{2}})^{2/3}\nonumber \\
 &  & +(\hat{X}_{A}^{\theta^{1}})^{2/3}(\hat{X}_{B}^{\phi^{3}})^{2/3}+(\hat{X}_{A}^{\theta^{1}})^{2/3}(\hat{X}_{B}^{\phi^{4}})^{2/3}\nonumber \\
 &  & +(\hat{X}_{A}^{\theta^{2}})^{2/3}(\hat{X}_{B}^{\phi^{1}})^{2/3}+(\hat{X}_{A}^{\theta^{2}})^{2/3}(\hat{X}_{B}^{\phi^{2}})^{2/3}\nonumber \\
 &  & +(\hat{X}_{A}^{\theta^{2}})^{2/3}(\hat{X}_{B}^{\phi^{3}})^{2/3}+(\hat{X}_{A}^{\theta^{2}})^{2/3}(\hat{X}_{B}^{\phi^{4}})^{2/3}\nonumber \\
 &  & +(\hat{X}_{A}^{\theta^{3}})^{2/3}(\hat{X}_{B}^{\phi^{1}})^{2/3}+(\hat{X}_{A}^{\theta^{3}})^{2/3}(\hat{X}_{B}^{\phi^{2}})^{2/3}\nonumber \\
 &  & +(\hat{X}_{A}^{\theta^{3}})^{2/3}(\hat{X}_{B}^{\phi^{3}})^{2/3}+(\hat{X}_{A}^{\theta^{3}})^{2/3}(\hat{X}_{B}^{\phi^{4}})^{2/3}\nonumber \\
 &  & +(\hat{X}_{A}^{\theta^{4}})^{2/3}(\hat{X}_{B}^{\phi^{1}})^{2/3}+(\hat{X}_{A}^{\theta^{4}})^{2/3}(\hat{X}_{B}^{\phi^{2}})^{2/3}\nonumber \\
 &  & +(\hat{X}_{A}^{\theta^{4}})^{2/3}(\hat{X}_{B}^{\phi^{3}})^{2/3}+(\hat{X}_{A}^{\theta^{4}})^{2/3}(\hat{X}_{B}^{\phi^{4}})^{2/3}\rangle\ ,\end{eqnarray}
where we denote sites $n=1,2$ as $A,B$ respectively, with phase
angles $\theta_{1}^{l}=\theta^{l}$ and $\theta_{2}^{l}=\phi^{l}$.
The averages of $\langle(\hat{X}_{A}^{\theta^{l}})^{\frac{1}{3}}(\hat{X}_{B}^{\phi^{l}})^{\frac{1}{3}}\rangle=0.433881cos(\phi^{l}-\theta^{l})$
and $\langle(\hat{X}_{A}^{\theta^{l}})^{\frac{2}{3}}(\hat{X}_{B}^{\phi^{l}})^{\frac{2}{3}}\rangle=0.42583$
are obtained following the methods given in the Appendix. We find
the value of the $LHS$ of Shchukin and Vogel's inequality is\begin{eqnarray}
LHS & = & \mid0.434881\{[cos(\phi^{1}-\theta^{1})-cos(\phi^{2}-\theta^{2})\nonumber \\
 &  & -cos(\phi^{3}-\theta^{3})-cos(\phi^{4}-\theta^{4})]\nonumber \\
 &  & +i[cos(\phi^{1}-\theta^{2})+cos(\phi^{2}-\theta^{1})\nonumber \\
 &  & -cos(\phi^{3}-\theta^{4})+cos(\phi^{4}-\theta^{3})]\nonumber \\
 &  & +j[cos(\phi^{1}-\theta^{3})+cos(\phi^{2}-\theta^{4})\nonumber \\
 &  & +cos(\phi^{3}-\theta^{1})-cos(\phi^{4}-\theta^{2})]\nonumber \\
 &  & +k[cos(\phi^{1}-\theta^{4})-cos(\phi^{4}-\theta^{3})\nonumber \\
 &  & +cos(\phi^{3}-\theta^{2})+cos(\phi^{4}-\theta^{1})]\}\mid^{2}\ ,\end{eqnarray}
using angles like $\theta^{1}=\phi^{1}=0,$ and $\theta^{2}=\theta^{3}=\theta^{4}=-\phi^{2}=-\phi^{3}=-\phi^{4}=-\pi/2$,
to get $LHS=\mid0.434881\times4\mid^{2}$ as all imaginary values
are $0$. There are many optimal choice of angles, but the maximum
values of the $LHS$ are same. The value of the $RHS$ is}

\textcolor{black}{\begin{eqnarray}
RHS & = & 16\langle(\hat{X}_{A}^{\theta^{1}})^{2/3}(\hat{X}_{B}^{\phi^{1}})^{2/3}\rangle\nonumber \\
 & = & 0.42583\times16\ ,\end{eqnarray}
so that the Bell ratio is the same as with the functional CFRD approach,
using a power law with $m=1/3$: }

\begin{eqnarray}
B_{2-SV(4)} & = & \frac{\mid0.434881\times4\mid^{2}}{0.42583\times16}\nonumber \\
 & = & \frac{\mid0.434881\times2\mid^{2}}{0.42583\times4}\nonumber \\
 & = & B_{2-CFRD}\ .\end{eqnarray}
\textcolor{black}{This is caused by the optimal value of the phases
corresponding effectively to only two observables per site, since
three of those phases have the same value.}

\subsubsection{Four site case}

\textcolor{black}{Next, we check mode number $N=4$, and $r=2$, for
states like $|\psi\rangle=1/\sqrt{2}(|0011\rangle+|1100\rangle)$.
In QM, the operators at each site in Shchukin and Vogel's theory are
\begin{eqnarray}
\hat{F}_{A} & = & (\hat{X}_{A}^{\theta^{1}})^{m}+i(\hat{X}_{A}^{\theta^{2}})^{m}+j(\hat{X}_{A}^{\theta^{3}})^{m}+k(\hat{X}_{A}^{\theta^{4}})^{m}\ ,\nonumber \\
\hat{F}_{B} & = & (\hat{X}_{B}^{\phi^{1}})^{m}+i(\hat{X}_{B}^{\mathbf{\phi^{2}}})^{m}+j(\hat{X}_{B}^{\phi^{3}})^{m}+k(\hat{X}_{B}^{\phi^{4}})^{m}\ ,\nonumber \\
\hat{F}_{C} & = & (\hat{X}_{C}^{\gamma^{1}})^{m}+i(\hat{X}_{C}^{\gamma^{2}})^{m}+j(\hat{X}_{C}^{\gamma^{3}})^{m}+k(\hat{X}_{C}^{\gamma^{4}})^{m}\ ,\nonumber \\
\hat{F}_{D} & = & (\hat{X}_{D}^{\varphi^{1}})^{m}+i(\hat{X}_{D}^{\varphi^{2}})^{m}+j(\hat{X}_{D}^{\varphi^{3}})^{m}+k(\hat{X}_{D}^{\varphi^{4}})^{m}\ ,\,\nonumber \\
 &  & \,\end{eqnarray}
where again $\theta_{1}^{l}=\theta^{l}$, $\theta_{2}^{l}=\phi^{l}$,
$\theta_{3}^{l}=\gamma^{l}$}, and \textcolor{black}{$\theta_{4}^{l}=\varphi{}^{l}$}.
After extensive algebra, we find that the largest Bell values with
optimal choice of angles are still the same as in the CFRD case using
$m=1/3$: $B_{4-SV(4)}=B_{4-CFRD}$. The reason for this identity,
as in the two-site case, is that with optimal settings we have effectively
only two measurements per site.

\subsection{Eight setting example}

\textcolor{black}{Finally, we can check the SV inequality with $8$
observables at each site. This approach uses octonions, and results
in a rather complex inequality \cite{Shchukin}. Therefore, we only
consider here the simplest two-site case $|\psi\rangle=c_{1}|0\rangle|1\rangle+c_{2}|1\rangle|0\rangle$,
$c_{1}=c_{2}=1/\sqrt{2}$. }

\textcolor{black}{We consider the fractional case $m=1/3$ as an example.
The LHS of the inequality is: \begin{eqnarray}
\mid\langle\hat{F}_{A}\hat{F}_{B}\rangle\mid^{2} & = & \mid0.434881\{[cos(\phi^{1}-\theta^{1})-cos(\phi^{2}-\theta^{2})\nonumber \\
 &  & -cos(\phi^{3}-\theta^{3})-cos(\phi^{4}-\theta^{4})\nonumber \\
 &  & -cos(\phi^{5}-\theta^{5})-cos(\phi^{6}-\theta^{6})\nonumber \\
 &  & -cos(\phi^{7}-\theta^{7})-cos(\phi^{8}-\theta^{8})]\nonumber \\
 &  & +i_{1}[cos(\phi^{1}-\theta^{2})+cos(\phi^{2}-\theta^{1})\nonumber \\
 &  & +cos(\phi^{3}-\theta^{4})-cos(\phi^{4}-\theta^{3})\nonumber \\
 &  & +cos(\phi^{5}-\theta^{6})-cos(\phi^{6}-\theta^{5})\nonumber \\
 &  & -cos(\phi^{7}-\theta^{8})+cos(\phi^{8}-\theta^{7})]\nonumber \\
 &  & +i_{2}[\cdots]+i_{3}[\cdots]+i_{4}[\cdots]+i_{5}[\cdots]\nonumber \\
 &  & +i_{6}[\cdots]+i_{7}[\cdots]\}\mid^{2}\ ,\end{eqnarray}
where $\hat{F}_{A}=(\hat{X}_{A}^{\theta^{1}})^{m}+i_{1}(\hat{X}_{A}^{\theta^{2}})^{m}+i_{2}(\hat{X}_{A}^{\theta^{3}})^{m}+i_{3}(\hat{X}_{A}^{\theta^{4}})^{m}+i_{4}(\hat{X}_{A}^{\theta^{5}})^{m}+i_{5}(\hat{X}_{A}^{\theta^{6}})^{m}+i_{6}(\hat{X}_{A}^{\theta^{7}})^{m}+i_{7}(\hat{X}_{A}^{\theta^{8}})^{m}$,
is an octonionic function of real observables, $(\hat{X}_{A}^{\theta^{l}})^{m},\ (l=1-8)$
at site A; similarly for site B, involving }$(\hat{X}_{B}^{\phi^{l}})^{m}$\textcolor{black}{.
The multiplication rules for the imaginary units of octonions were
listed in \cite{Shchukin}. One of the optimal choices of angles still
keeps the same pattern as in two-sites case: $\theta^{1}=\phi^{1}=0,$
and $\theta^{l\neq1}=-\phi^{l\neq1}=-\pi/2$, so that we obtain $LHS=\mid0.434881\times8\mid^{2}$,
and \begin{eqnarray}
RHS & = & 64\langle(\hat{X}_{A}^{\theta^{1}})^{2/3}(\hat{X}_{B}^{\phi^{1}})^{2/3}\rangle\ .\end{eqnarray}
Therefore \begin{eqnarray}
B_{2-SV(8)}= & \frac{\mid0.434881\times8\mid^{2}}{0.4258\times64} & =B_{2-CFRD}\ .\end{eqnarray}
}

\textcolor{black}{For $N=4$ modes, the optimal result of the SV inequality
with $8$ observables at each site is also the same as for the functional
CFRD fractional moment inequality, for the particular state $|\psi\rangle=1/\sqrt{2}(|0011\rangle+|1100\rangle)$.}

In summary, in the cases investigated, the results are the same as
those found with the simpler CFRD approach, using fractional moments.
The reason is that the optimal values can be found with only 2 effective
settings per site with the types of states we consider. However, this
situation may change with more general input states or functional
transformations. We leave this as an open question for future work.

\section{Sensitivity to loss and mixing}

The value of the Bell observable $B_{N}$ or $S_{N}$ increases with
the number of sites $N$, and this is suggestive of a strategy that
may allow genuine loophole-free violations of LR. However, it may
be argued that since increasing the number of sites will increase
the number of detectors required, there will be no advantage. Only
careful calculation of the Bell observable $B_{N}$ including the
detection efficiency $\eta$ can determine whether the strategy is
advantageous for loophole-free violation of Bell inequalities.

\subsection{Model of loss and impurity}

Loss is modelled using a beam splitter as follows. The field modes
$\hat{a}_{in}$ at each site are independently coupled to a second
mode $\hat{a}_{v}$, assumed to be in a vacuum state. Bosons are lost
from the field into the vacuum mode, the strength of coupling determining
the rate of loss. For each mode, the beam splitter model gives the
final detected and vacuum mode in terms of the inputs $\hat{a}_{in}$
and $\hat{a}_{v}$ \begin{eqnarray}
\hat{a}_{out} & = & \sqrt{\eta}\hat{a}_{in}+\sqrt{1-\eta}\hat{a}_{v}\ ,\nonumber \\
\hat{a}_{v,out} & = & \sqrt{1-\eta}\hat{a}_{in}-\sqrt{\eta}\hat{a}_{v}\ ,\end{eqnarray}
where $\eta$ is the efficiency, the probability for detecting a boson,
given that one is incident. One can rewrite these relations to obtain
\begin{eqnarray}
\hat{a}_{in} & = & \sqrt{\eta}\hat{a}_{out}+\sqrt{1-\eta}\hat{a}_{v,out}\ ,\nonumber \\
\hat{a}_{v} & = & \sqrt{1-\eta}\hat{a}_{out}-\sqrt{\eta}\hat{a}_{v,out}\ .\end{eqnarray}
Since we only measure {}``$\hat{a}_{out}$'', not {}``$\hat{a}_{v,out}$'',
we need to trace over all the unmeasured vacuum output modes to obtain
the final density operator for the detected modes after loss. 

Thus, given a total density matrix of $\hat{\rho}$, we only measure
a reduced density matrix $\hat{\rho}_{out}$, where:\begin{eqnarray}
\hat{\rho}_{out} & = & \mathrm{Tr}_{v,out}\{\hat{\rho}\}\nonumber \\
 & = & \sum_{k}\langle\psi_{k}|_{v,out}\hat{\rho}|\psi_{k}\rangle_{v,out}\end{eqnarray}

\textcolor{black}{We also examine the effect of impurity, by considering
an input state $\hat{\rho}'=p|\psi\rangle\langle\psi|+(1-p)\hat{\rho}_{mix}$,
where $\hat{\rho}_{mix}$ is the mixed state obtained by complete
decoherence in the occupation-number basis, i.e. \begin{eqnarray}
\hat{\rho}_{mix} & = & |c_{1}|^{2}|0\rangle^{\otimes r}|1\rangle^{\otimes(N-r)}\langle0|{}^{\otimes r}\langle1|^{\otimes(N-r)}+\nonumber \\
 &  & +|c_{2}|^{2}|1\rangle^{\otimes r}|0\rangle^{\otimes(N-r)}\langle1|{}^{\otimes r}\langle0|^{\otimes(N-r)}\,\,.\end{eqnarray}
Here $p$ is the probability the system is in the original pure state
\eqref{eq:GHZ}. While this is not the most general model of state
impurity, it is a relatively simple one that allows us to carry out
indicative calculations of the effects of impure input states.}

\subsection{Three-mode MABK example}

We can transform the input state for $N=3$ and $r=1$, $|\psi\rangle=\frac{1}{\sqrt{2}}(|100\rangle+|011\rangle)|000\rangle$,
for example, where $|000\rangle$ represents the vacuum fields, into
the output after loss. We denote the three field mode operators involved
as $\hat{a}$, $\hat{b}$, $\hat{c}$ for clarity. Thus in this case,
\begin{eqnarray}
|out\rangle & = & [\hat{a}_{in}^{\dagger}|000\rangle|000\rangle+\hat{b}_{in}^{\dagger}\hat{c}_{in}^{\dagger}|000\rangle|000\rangle]/\sqrt{2}\nonumber \\
 & = & (\sqrt{\eta}|100\rangle_{out}|000\rangle_{v,out}+\nonumber \\
 &  & \sqrt{1-\eta}|000\rangle_{out}|100\rangle_{v,out}+\eta|011\rangle_{out}|000\rangle_{v,out}\nonumber \\
 &  & +(1-\eta)|000\rangle_{out}|011\rangle_{v,out}+\nonumber \\
 &  & \sqrt{1-\eta}\sqrt{\eta}(|001\rangle_{out}|010\rangle_{v,out}\nonumber \\
 &  & +|010\rangle_{out}|001\rangle_{v,out})/\sqrt{2}\ .\end{eqnarray}
Next, we need to trace over the vacuum output modes to obtain a reduced
density matrix $\hat{\rho}_{out}$, where: \begin{eqnarray}
\hat{\rho}_{out} & = & \mathrm{Tr}_{v,out}\{\hat{\rho}\}\nonumber \\
 & = & \sum_{i,j,k}\langle ijk|_{v,out}\hat{\rho}|ijk\rangle_{v,out}\nonumber \\
 & = & \frac{1}{2}\{[\sqrt{\eta}|100\rangle+\eta|011\rangle][\sqrt{\eta}\langle100|+\eta\langle011|]+\nonumber \\
 &  & (1-\eta)|000\rangle\langle000|+\eta(1-\eta)|001\rangle\langle001|\nonumber \\
 &  & +\eta(1-\eta)|010\rangle\langle010|+\nonumber \\
 &  & +(1-\eta)^{2}|000\rangle\langle000|\}\ .\end{eqnarray}
Proceeding, the probability distributions are evaluated. Thus, in
this three-mode example: 

\begin{eqnarray}
P(x_{\theta},x_{\phi},x_{\gamma}) & = & \langle x_{\theta}|\langle x_{\phi}|\langle x_{\gamma}|\hat{\rho}_{out}|x_{\theta}\rangle|x_{\phi}\rangle|x_{\gamma}\rangle\nonumber \\
 & = & \frac{1}{2}(\frac{2}{\pi})^{3/2}e^{-2(x_{\theta}^{2}+x_{\phi}^{2}+x_{\gamma}^{2})}\times\nonumber \\
 &  & \times\left[\right.\mid(2\sqrt{\eta}e^{i\theta}x_{\theta}+4\eta e^{i\phi}e^{i\gamma}x_{\phi}x_{\gamma})\mid^{2}+\nonumber \\
 &  & +4\eta(1-\eta)\times\left\{ \right.\left\{ \mid x_{\gamma}\mid^{2}+\mid x_{\phi}\mid^{2}\right\} \nonumber \\
 &  & +[1-\eta+(1-\eta)^{2}]\left.\right]\,\,,\end{eqnarray}
and the integrals give for example 

\begin{eqnarray}
\langle\hat{X}_{A}^{\theta,bin}\hat{X}_{B}^{\phi,bin}\hat{X}_{C}^{\gamma,bin}\rangle & = & \left(\frac{2\eta}{\pi}\right)^{\frac{3}{2}}\cos(\gamma+\phi-\theta).\end{eqnarray}

Examining the MABK Bell inequalities defined by (\ref{eq:MK}), we
finally obtain, using the {}``orthogonal'' choice of measurements
given for (\ref{eq:anlgeorth}), the result: \begin{eqnarray}
\mid\langle\hat{a}_{bin}^{\dagger}\hat{b}_{bin}\hat{c}_{bin}\rangle\mid & = & 4\left(\frac{2\eta}{\pi}\right)^{\frac{3}{2}}\ .\end{eqnarray}
Alternatively, for the \textcolor{black}{measurement }angles \textcolor{black}{$\theta=0,\ \phi=-\pi/6,\ \gamma=-2\pi/6,\ \theta'=\pi/2,\ \phi'=-4\pi/6,\ \gamma'=-5\pi/6,$
we can rotate the $LHS$ moment, but so that again}

\begin{equation}
\mid\langle\hat{a}_{bin}^{\dagger}\hat{b}_{bin}\hat{c}_{bin}\rangle\mid=4\left(\frac{2\eta}{\pi}\right)^{\frac{3}{2}}\ .\end{equation}

\subsection{Higher N results for MABK inequalities}

\textcolor{black}{Continuing in this fashion for higher $N,$ we find}\textcolor{red}{{}
}that the values of the Bell observable including the effect of detection
inefficiencies with the optimal choice of angles is \begin{equation}
S_{N}=\frac{1}{\sqrt{2}}\left(\frac{4\eta}{\pi}\right)^{\frac{N}{2}}\ ,\end{equation}
which implies a critical minimum efficiency \begin{equation}
\eta_{min}=2^{(1-2N)/N}\pi\end{equation}
in order to violate the MABK inequalities (\ref{eq:MK}).

\textcolor{black}{We can also examine the effect of impurity or noise,
by considering a state $\hat{\rho}'=p|\psi\rangle\langle\psi|+(1-p)\hat{\rho}_{mix}$,
as described above.$\ $ The effect of this type of `noise' on the
MABK inequality is to give \begin{equation}
S_{N}(p,\eta)=\frac{\sqrt{2}}{2}\left(\frac{4\eta p^{2}}{\pi}\right)^{\frac{N}{2}}\,\,.\end{equation}
This result agrees with that obtained by Acin et al \cite{Acin }
for the case $\eta=1$. }

\begin{figure}[h]
\textcolor{black}{\includegraphics[width=1\columnwidth,height=5.8cm]{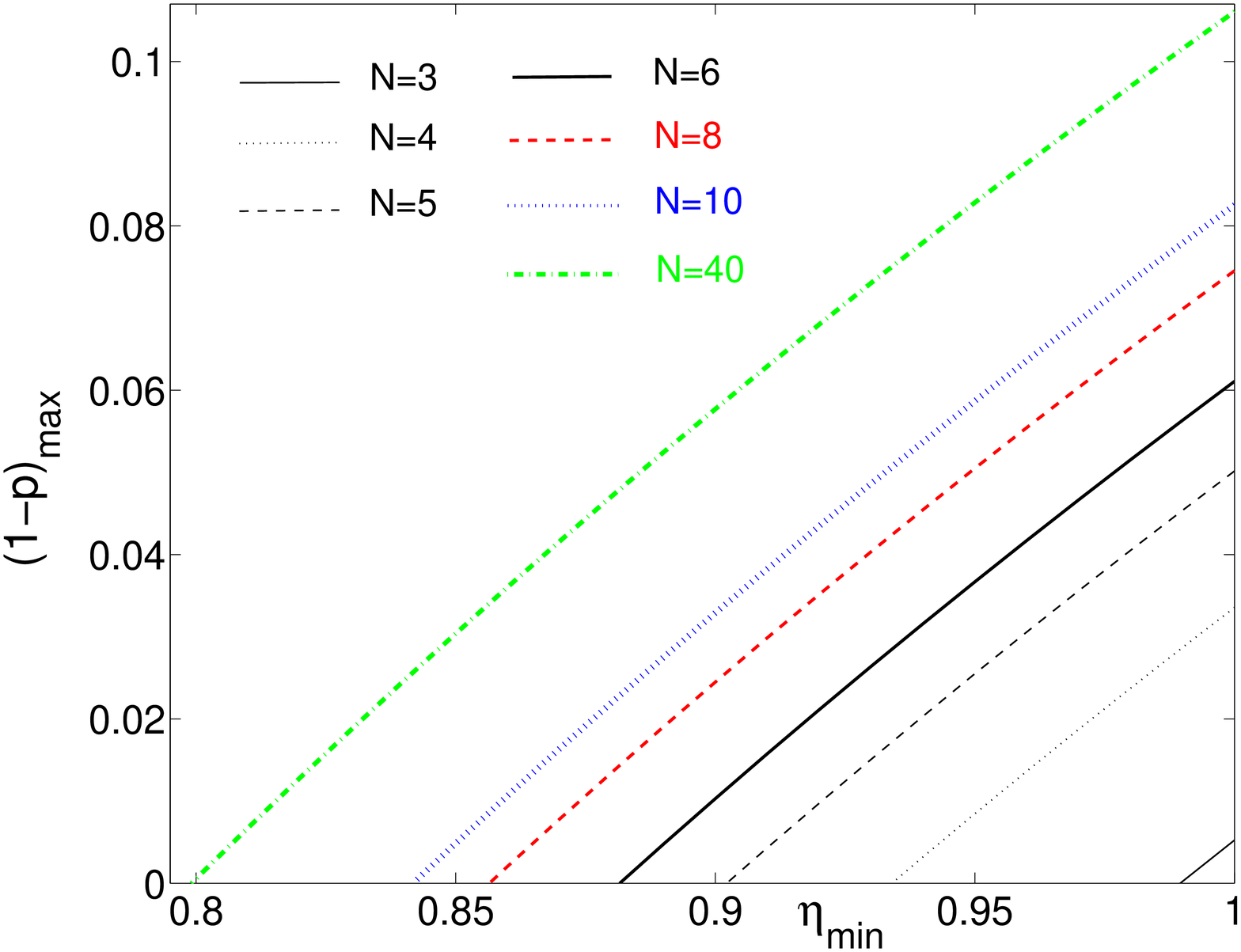}}

\textcolor{black}{\caption{\textcolor{black}{(Color online) The boundary of maximum noise $ $$(1-p)_{max}$
and minimum detection efficiency $\eta_{min}$ to violate the}~\textcolor{black}{MABK
Bell inequality with binned outcomes. }Different $N$ is distinguished
by the line style.\textcolor{black}{\label{fig:boundary of loss and noise-MK}}}
}
\end{figure}

\textcolor{black}{For lower $N$, the strategy of binning and using
the MABK inequalities shows an advantage, by allowing a violation
of LHV theories for $N=3,\ 4,\ 5$. However, very high efficiencies,
$\eta>0.99$, $0.93$, $0.90$ respectively, even for a pure state
$p=1$, are required, as shown in Fig. \ref{fig:boundary of loss and noise-MK}.
While high detection efficiencies are feasible for homodyne detection,
these efficiency values are still quite challenging once generation
losses are also taken into account. In view of this, the high requirement
for $\eta_{min}$ for the case $N=3$ may be prohibitive. }

\textcolor{black}{}%

\textcolor{black}{}%
\begin{figure}[h]
\textcolor{black}{\includegraphics[width=1\columnwidth,height=5.8cm]{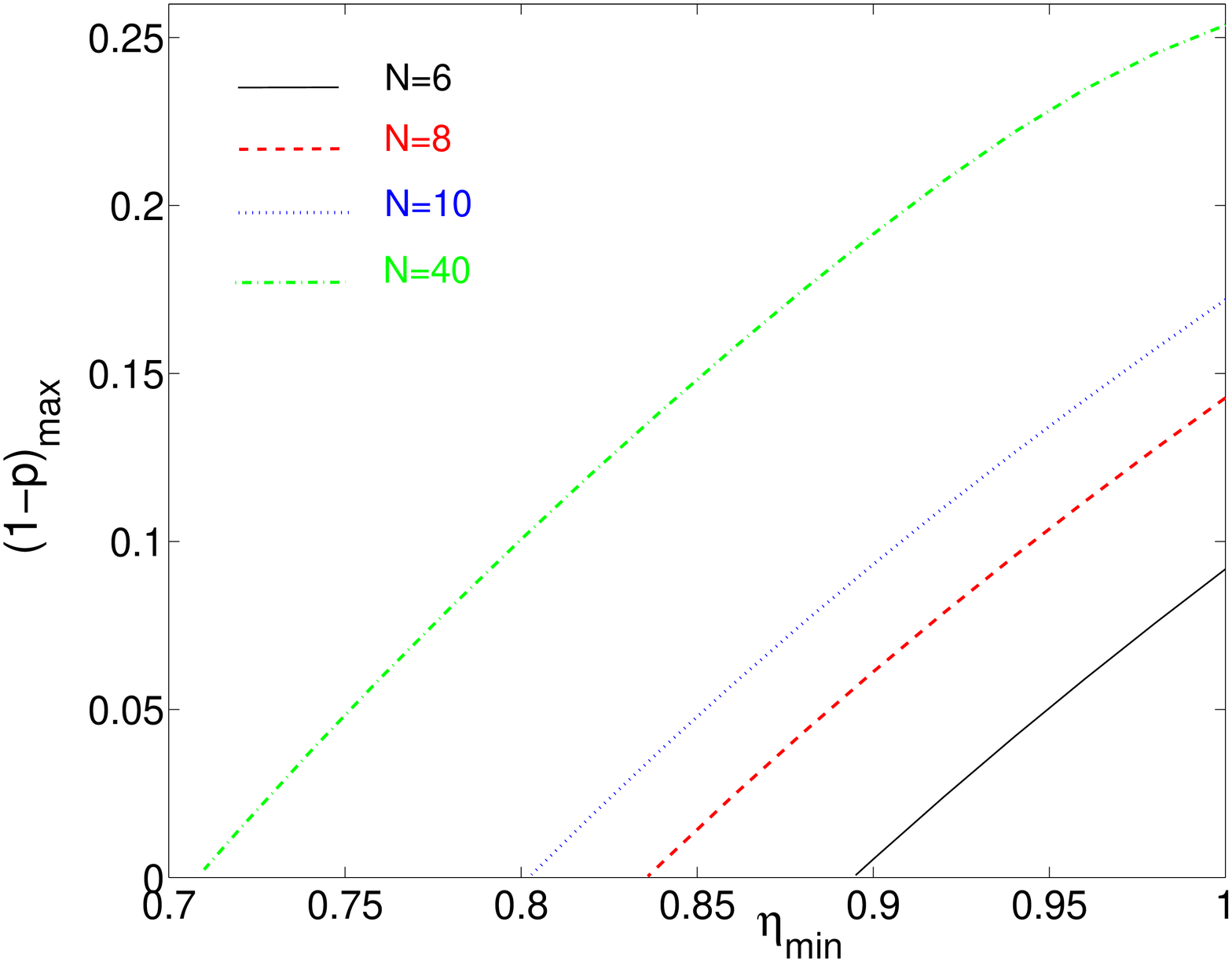}}

\textcolor{black}{\caption{\textcolor{black}{(Color online) The boundary of maximum noise $ $$(1-p)_{max}$
and minimum detection efficiency $\eta_{min}$ to violate the functional
CFRD Bell inequality with observables as the optimal function. }Different
$N$ is distinguished by the line style.\textcolor{black}{{} \label{fig:boundary of loss and noise}}}
}
\end{figure}

\subsection{Loss and noise with functional inequalities}

\textcolor{black}{This approach is also applied to enable a prediction
of the effect of loss and noise on the functional inequalities of
Section IV, and the results are plotted in Fig. \ref{fig:boundary of loss and noise}
for the optimum function.} Here, the parameter $\varepsilon_{N}$
is changed to $\varepsilon_{N}(\eta)$ for the optimum function. For
$N$ even we find that\begin{eqnarray}
\varepsilon_{N}(\eta) & = & \frac{2\eta\varepsilon_{N}}{2\eta+(1-\eta)\varepsilon_{N}}\ ,\nonumber \\
B_{N} & = & 2^{N-2}[\frac{2(I^{+})^{4}(\eta p)^{2}}{\pi I^{o}C}]^{\frac{N}{2}}\ ,\label{eq:Even Bell}\end{eqnarray}
where $\varepsilon_{N}$ is defined as before, and $C=\eta I+(1-\eta)I^{o}$.
For the case of odd $N$ the relevant integral equations change, giving\textcolor{black}{{}
a modified (and slightly reduced) Bell variable $B'_{N}$, where:} 

\begin{eqnarray}
\varepsilon'_{N}(\eta) & = & \varepsilon_{N}(\eta)\frac{N\varepsilon_{N}^{+}(\eta)-\varepsilon_{N}(\eta)\varepsilon_{N}^{-}/\varepsilon_{N}}{N\varepsilon_{N}^{+}(\eta)+\varepsilon_{N}^{2}(\eta)\varepsilon_{N}^{-}/\varepsilon_{N}^{2}}\ ,\nonumber \\
B'_{N} & = & \frac{2\sqrt{I^{o}C}}{I^{o}+C}B_{N}\ .\label{eq:Odd Bell}\end{eqnarray}
\textcolor{black}{Here $\varepsilon_{N}^{+}(\eta)=\varepsilon_{N}(\eta)+4$,
and $B_{N}$ is defined as in Eq (\ref{eq:Even Bell}).}

For pure state, in the case of $r=N/2$, $N$ is even, we can give
the threshold efficiency requirement analytically since all integrals
are independent of $N$, where 

\begin{equation}
\eta_{crit}=\frac{[(4-\varepsilon_{N})+\sqrt{(4-\varepsilon_{N})^{2}+4\varepsilon_{N}^{2}}]I_{0}I\pi}{2^{6-4/N}(I^{+})^{4}}\ .\end{equation}
This reduces at large $N$ to an asymptotic value of $\eta_{\infty}=0.69$,
thus dramatically reducing the required detection efficiency. The
violation is less sensitive to detector inefficiency in the macroscopic,
large $N$ limit. 

\textcolor{black}{Compared with the MABK binning approach, the robustness
of the functional inequality against noise and inefficiency is apparent
as shown in Fig. \ref{fig:boundary of loss and noise}. For $N>7$,
the functional CV inequalities used with an }\textit{\textcolor{black}{optimal}}\textcolor{black}{{}
function shows clear advantages, allowing violation of LHV at much
lower efficiencies and larger maximum noise. For example, for $N=40$,
we can get Bell violation with $\eta_{crit}\sim70.8\%$ for pure state,
or $p_{max}\sim0.25$ for ideal detection efficiency. $\ $The potentially
feasible $\eta_{crit}\sim80\%$ requires $N\approx10$ for a pure
state if the functional optimization is used. By comparison, the binned
case requires $N\sim40$, and the original CFRD cannot reach this
even with $N\rightarrow\infty$.}

\section{Conclusion}

\textcolor{black}{In summary, we have developed a functional approach
to tests of quantum nonlocality, and showed how this gives an exceptional
degree of robustness against losses and noise. For continuous variable
measurements on GHZ states, this approach gives distinct advantages
when compared with either binning methods or the moment-based approach
of Cavalcanti }\textcolor{black}{\emph{et al.~}}\textcolor{black}{(CFRD)
\cite{cvbell2}. We have also compared the class of inequalities derived
by CFRD}\textcolor{black}{\emph{ }}\textcolor{black}{with the generalizations
given by Shchukin and Vogel (SV) for more observables at each site
but found no better violations with the photon-number correlated states
introduced here. }

\textcolor{black}{However, there are still many open questions. The
first is applicable even to the CFRD result: we do not know yet if
there are any other quantum states that can give better violations
than those given here for the GHZ-type states. Another question is
whether there are other states for which the multiple-observable SV
inequalities provide an advantage over the CFRD. One may also ask
if there are yet more general continuous variable inequalities than
either the CFRD or SV results.}

\textcolor{black}{Further applications of these ideas may include
application of the functional approach to tests of other forms of
nonlocality --- e.g, entanglement and EPR-steering \cite{prasteerepr}.
It seems possible that a functional optimization will allow more robust
implementations of these tests as well.}

\section*{ACKNOWLEDGEMENT}

\textcolor{black}{We wish to acknowledge funding for this project
from the Australian Research Council through a Discovery grant and
the Australian Research Council Center of Excellence for Quantum-Atom
Optics.}

\section*{{\normalsize Appendix I: Quadrature Wavefunctions }}

Calculations with the functional transformations used here rely on
the fact that quadrature variables have the same wave-functions as
the quantum harmonic oscillator. We first recall that the harmonic
oscillator Hamiltonian with unit mass and frequency $\omega$ can
be written:\begin{equation}
\hat{H}=\frac{1}{2}(\hat{p}^{2}+\omega^{2}\hat{x}^{2})\,\,.\end{equation}
Here $\hat{p}$ is the momentum and $\hat{x}$ is the position operator,
with commutators $\left[\hat{x},\hat{p}\right]=i\hbar$, and creation
and annihilation operators defined as: \begin{eqnarray}
\hat{a} & = & (2\hbar\omega)^{-1/2}(\omega\hat{q}+i\hat{p})\ ,\nonumber \\
\hat{a}^{\dagger} & = & (2\hbar\omega)^{-1/2}(\omega\hat{q}-i\hat{p})\ .\end{eqnarray}
For simplicity we assume that $\omega=1$ and $\hbar=1/2$, which
corresponds to the convention used throughout the paper: \begin{eqnarray}
\hat{x} & = & (\hat{a}^{\dagger}+\hat{a})/2\ ,\nonumber \\
\hat{p} & = & -i(\hat{a}-\hat{a}^{\dagger})/2\ .\end{eqnarray}

\subsection{Wavefunction for quadrature variables}

In this standard form, the harmonic oscillator wavefunction \cite{Louisell-1}
for the $n^{th}$ energy eigenstate is:\begin{equation}
\langle x|n\rangle=(\frac{2}{\pi2^{2n-1}\left[n!\right]^{2}})^{1/4}H_{n}(\sqrt{2}x)e^{-x^{2}}\ ,\label{eq:q-wavefunction}\end{equation}
where $H_{n}(x)$ is the $n^{th}$ Hermite polynomial. Thus, for example:\begin{eqnarray}
\langle x|0\rangle & = & \left(\frac{2}{\pi}\right)^{1/4}e^{-x^{2}}\ ,\nonumber \\
\langle x|1\rangle & = & 2x\left(\frac{2}{\pi}\right)^{1/4}e^{-x^{2}}\ .\end{eqnarray}

We also need the complementary (momentum) wavefunctions:\begin{equation}
\langle p|\psi\rangle=\frac{1}{\sqrt{\pi}}\int_{-\infty}^{\infty}exp(-2ip'x')\langle x'|\psi\rangle dx'\ ,\end{equation}
 Again, using normalized quadrature variables, one finds that:\textcolor{black}{\begin{eqnarray}
\langle p|0\rangle & = & \left(\frac{2}{\pi}\right)^{1/4}e^{-p^{2}}\ ,\nonumber \\
\langle p|1\rangle & = & -2ip\left(\frac{2}{\pi}\right)^{1/4}e^{-p^{2}}\ .\end{eqnarray}
}

\subsection{General rotation}

\textcolor{black}{Next, we consider the case of a general rotation
of the quadrature phase by defining\[
\hat{X}^{\theta}=\hat{x}\cos\theta+\hat{p}\sin\theta=(\hat{a}e^{-i\theta}+\hat{a}^{\dagger}e^{i\theta})/2\ ,\]
}Hence we find that\textcolor{black}{\begin{eqnarray}
\langle X^{\theta}|0\rangle & = & \left(\frac{2}{\pi}\right)^{1/4}e^{-(X^{\theta})^{2}}\ ,\nonumber \\
\langle X^{\theta}|1\rangle & = & 2e^{-i\theta}\left(\frac{2}{\pi}\right)^{1/4}X^{\theta}e^{-(X^{\theta})^{2}}\ ,\end{eqnarray}
and $p$ corresponds to $\theta=\pi/2$.}

\subsection{Joint probabilities}

\textcolor{black}{To carry out moment calculations, it is necessary
to have joint probabilities. As an example of this calculation, we
consider a two-mode state $\hat{\rho}=\left|\Psi\right\rangle \left\langle \Psi\right|$,
where: \begin{equation}
\left|\Psi\right\rangle =\frac{1}{\sqrt{2}}\left[\left|0,1\right\rangle +\left|1,0\right\rangle \right]\ .\end{equation}
}

\textcolor{black}{Carrying out the calculation of the joint probability,
and defining $\left|\mathbf{X}\right|^{2}=(X^{\theta})^{2}+(X^{\phi})^{2}$,
we find that:}

\begin{eqnarray}
{\color{black}P(X^{\theta},X^{\phi})} & = & {\color{black}\langle X^{\theta}|\langle X^{\phi}|\hat{\rho}|X^{\theta}\rangle|X^{\phi}\rangle}\nonumber \\
 & {\color{black}=} & {\color{black}\frac{1}{2}|\langle X^{\theta}|0\rangle\langle X^{\phi}|1\rangle+\langle X^{\theta}|1\rangle\langle X^{\phi}|0\rangle|^{2}}\nonumber \\
 & {\color{black}=} & {\color{black}\left(\frac{4}{\pi}\right)|e^{-\left|\mathbf{X}\right|^{2}}(e^{i\phi}X^{\phi}+e^{i\theta}X^{\theta})|^{2}}\nonumber \\
 & = & \left(\frac{4}{\pi}\right)e^{-2\left|\mathbf{X}\right|^{2}}\times\nonumber \\
 &  & \times[\left|\mathbf{X}\right|^{2}+2X^{\theta}X^{\phi}cos(\theta-\phi)]\ \label{eq:probability}\end{eqnarray}

\subsection{Binned moment calculation}

As an example of the application of the joint probability, consider
a binned moment calculation for the case $N=2$, using an initial
Bell state of the form given above. Then we \textcolor{black}{obtain:}

\textcolor{red}{\begin{eqnarray}
{\color{black}\langle\hat{X}_{A}^{\theta,bin}\hat{X}_{B}^{\phi,bin}\rangle} & {\color{black}=} & {\color{black}\int_{X^{\theta}=0}^{\infty}\int_{X^{\phi}=0}^{\infty}P(X^{\theta},X^{\phi})d^{2}\mathbf{X}}\nonumber \\
{\color{black}{\color{red}}} & {\color{black}{\color{red}}} & {\color{black}-\int_{X^{\theta}=0}^{\infty}\int_{X^{\phi}=-\infty}^{0}P(X^{\theta},X^{\phi})d^{2}\mathbf{X}}\nonumber \\
{\color{black}{\color{red}}} & {\color{black}{\color{red}}} & {\color{black}+\int_{X^{\theta}=-\infty}^{0}\int_{X^{\phi}=-\infty}^{0}P(X^{\theta},X^{\phi})d^{2}\mathbf{X}}\nonumber \\
{\color{black}{\color{red}}} & {\color{black}{\color{red}}} & {\color{black}-\int_{X^{\theta}=-\infty}^{0}\int_{X^{\phi}=0}^{\infty}P(X^{\theta},X^{\phi})d^{2}\mathbf{X}}\nonumber \\
{\color{black}{\color{red}}} & {\color{black}=} & {\color{black}\left(\frac{2}{\pi}\right)\cos(\phi-\theta)\ .}\end{eqnarray}
}

\subsection{Functional moment calculation}

Here we show how to calculate the expectation value of products of
functional moments using the joint probabilities. For the case $N=2$,
using the same state given above, the value of the LHS is

\begin{eqnarray}
LHS & = & \left|\left\langle \prod_{n=1}^{2}\left[f_{n}(\hat{X}_{n}^{\theta})+ig_{n}(\hat{X}_{n}^{\theta'})\right]\right\rangle \right|^{2}\,\,.\end{eqnarray}
This can be evaluated by the joint probability shown in Eq. (\ref{eq:probability}),
and will include terms like:\begin{multline}
{\color{black}\langle f_{1}(\hat{X}_{1}^{\theta})f_{2}(\hat{X}_{2}^{\phi})\rangle}=\left(\frac{4}{\pi}\right)\int\int e^{-2\left|\mathbf{X}\right|^{2}}f_{1}f_{2}[\left|\mathbf{X}\right|^{2}\\
+2X^{\theta}X^{\phi}cos(\theta-\phi)]d^{2}\mathbf{X}\,\,.\label{eq:exp_val_f1_f2}\end{multline}

To show this more clearly, we give an example of the observables as
fractional order moments $(\hat{X}^{\theta})^{m}$ with corresponding
outcomes $|x^{\theta}|^{m}sign(x^{\theta})$. Replacing $f_{1}$ and
$f_{2}$ in the above integral with $(\hat{X}_{A}^{\theta})^{m}$
and $(\hat{X}_{B}^{\theta})^{m}$, we obtain that \begin{eqnarray}
\langle(\hat{X}_{A}^{\theta})^{\frac{1}{3}}(\hat{X}_{B}^{\phi})^{\frac{1}{3}}\rangle & = & 0.433881cos(\phi-\theta)\ ,\nonumber \\
\langle(\hat{X}_{A}^{\theta})^{\frac{2}{3}}(\hat{X}_{B}^{\phi})^{\frac{2}{3}}\rangle & = & 0.42583\ \end{eqnarray}
for $m=1/3$ and $m=2/3$. 

For convenience, we assume the functions $f_{n}$ and $g_{n}$ are
odd, and thereby only the cross term in \eqref{eq:exp_val_f1_f2}
has a nonzero contribution. The value of $LHS$ can be maximized by
using orthogonal angles so that $cos(\theta-\phi)=\pm1$: \begin{eqnarray}
LHS & = & \left|\left\langle \frac{\prod_{n=1}^{2}(f_{n}+g_{n})+\prod_{n=1}^{2}(f_{n}-g_{n})}{2}\right\rangle \right|^{2}\nonumber \\
 & = & \left|\left(\frac{1}{\pi}\right)\left[\prod_{n=1}^{2}e^{-2x^{2}}xf_{n}^{+}dx+\prod_{n=1}^{2}e^{-2x^{2}}xf_{n}^{-}dx\right]\right|^{2}\nonumber \\
 & = & \left|\left(\frac{1}{\pi}\right)\left[\prod_{n=1}^{2}I_{n}^{+}+\prod_{n=1}^{2}I_{n}^{-}\right]\right|^{2}\ ,\end{eqnarray}
\textcolor{black}{where $f_{n}^{\pm}=f_{n}\pm g_{n}$, and define
integrals for $x$: $I_{n}^{\pm}=2\int e^{-2x^{2}}xf_{n}^{\pm}dx$.}

While the value of RHS 

\begin{eqnarray}
RHS & = & \langle\prod_{n=1}^{2}\left[f_{n}(\hat{X}_{n}^{\theta})^{2}+g_{n}(\hat{X}_{n}^{\theta'})^{2}\right]\rangle\nonumber \\
 & = & \langle\prod_{n=1}^{2}\frac{(\hat{f}_{n}+\hat{g}_{n})^{2}+(\hat{f}_{n}-\hat{g}_{n})^{2}}{2}\rangle\end{eqnarray}
is invariant with angles as it only related to the square items in
the probability of Eq. (\ref{eq:probability}). Define \textcolor{black}{$I_{n}=4\int x^{2}e^{-2x^{2}}[(f_{n}^{+})^{2}+(f_{n}^{-})^{2})]dx$,
and $I_{n}^{o}=\int e^{-2x^{2}}[(f_{n}^{+})^{2}+(f_{n}^{-})^{2})]dx$,
then we can obtain\begin{eqnarray}
RHS & = & \frac{1}{4\pi}[\prod_{n=1}^{r}I_{n}\prod_{k=r+1}^{N}I_{n}^{o}+\prod_{n=1}^{r}I_{n}^{o}\prod_{k=r+1}^{N}I_{n}],\nonumber \\
 &  & \,\end{eqnarray}
}where $r=1$, $N=2$.


\begin{thebibliography}{34}
\bibitem{key-1}Bearing in mind these measurements can be noncompatible,
the LHS is simply an abbreviation for the expansion in terms of observable
moments, eg $\bigl\{\langle x_{1}x_{2}\rangle-\langle p_{1}p_{2}\rangle\bigl\}^{2}+\bigl\{\langle x_{1}p_{2}\rangle+\langle x_{2}p_{1}\rangle\bigl\}^{2}$
when $N=2$.
\end{thebibliography}
\end{document}